\pdfoutput=1
\RequirePackage{amsmath}
\documentclass[twocolumn]{svjour3}          
\smartqed  

\usepackage{amssymb}
\usepackage{amsfonts}

\usepackage{graphicx}
\usepackage{mathptmx}      

\usepackage[svgnames]{xcolor}
\usepackage{tikz}
\usepackage{pgfplots}

\usepackage{hyphenat}   

\usepackage{contour}    

\usepackage[hidelinks]{hyperref}
\hypersetup{
  colorlinks   = true, 
  urlcolor     = DarkOrchid, 
  linkcolor    = MediumBlue, 
  citecolor   = MediumBlue 
}

\DeclareMathAlphabet{\mathcal}{OMS}{cmsy}{m}{n}   

\journalname{Shock Waves}

\pgfplotsset{compat=1.12}

\newcommand{\mdqm}{$\mathrm{MD_4M}$}
\newcommand{\mmg}{{Mmg}}
\newcommand{\testA}{\textsf{Test} $\mathcal{D}$}
\newcommand{\testB}{\textsf{Test} $\mathcal{NI}$}
\newcommand{\testC}{\textsf{Test} $\mathcal{NC}$}

\newcommand{\CC} {\mathcal{C}}
\newcommand{\dCC} {\partial\mathcal{C}}
\newcommand{\Et}{E^\mathrm{t}}

\newcommand{\xx} {\mathbf{x}}
\newcommand{\uu} {\mathsf{u}}
\newcommand{\ff} {\mathbf{f}}
\newcommand{\mm} {\mathbf{m}}
\newcommand{\nn} {\mathbf{n}}
\newcommand{\vv} {\mathbf{v}}

\newcommand{\Ci} {\mathcal{C}_i}
\newcommand{\ui} {\mathsf{u}_i}

\newcommand{\etab}{\contour[2]{black}{$\eta$}}
\newcommand{\xib }{\contour[2]{black}{$\xi$}}
\newcommand{\KK} {\mathcal{K}}

\newcommand{\dxB} {\Delta \mathbf{x}_\mathsf{B}^{n+1}}

\newcommand{\tn}{^n}
\newcommand{\tnn}{^{n+1}}
\newcommand{\tns}{^{n+}}

\newcommand{\crit}{_\mathrm{c}}
\newcommand{\Tb}{T_\mathrm{b}}

\DeclareRobustCommand\line[1]{%
  \begin{tikzpicture}    
    \path (0pt, -0.5ex) -- (0pt,0pt);                                                    
    \draw[#1] (0pt,0pt) -- (10pt,0pt);                                       
  \end{tikzpicture}%
}

\DeclareRobustCommand\markCaption[1]{%
  \begin{tikzpicture}
\begin{axis}[hide axis, scale only axis, height=2ex, xmin=0, xmax=0.001, ymin=0, ymax=0.001]
\addplot[#1] coordinates {(0,0)};
\end{axis}                                       
  \end{tikzpicture}%
} 

\DeclareRobustCommand\square[1]{%
  \begin{tikzpicture}                                                           
    \fill[#1] (0pt,0pt) rectangle (5pt,4pt);                                       
  \end{tikzpicture}%
}

\newcommand{\orcidlogo}[1]{\href{https://orcid.org/#1}{\includegraphics[scale=0.5]{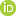}}}

\graphicspath{{./../3.0_Figures_def/}}

\begin{document}

\title{An Adaptive ALE Scheme for Non-Ideal Compressible-Fluid Dynamics over Dynamic Unstructured Meshes}


\author{Barbara Re \orcidlogo{0000-0003-1692-2053}  \and
        Alberto Guardone
}
\authorrunning{Barbara Re \and Alberto Guardone}

\institute{B. Re$^\star$ \and A. Guardone
           \at
           Department of Aerospace Science and Technology, Politecnico di Milano, 20156 Milano, Italy \\
              \email{barbara.re@polimi.it,  alberto.guardone@polimi.it} \\ 
              $^\star$\emph{Present address:} Institute of Mathematics,
              University of Z\"{u}rich, 8057 Z\"{u}rich, Switzerland  
}

\date{Received: 13 October 2017 / Revised: 4 June 2018/ Accepted: 6 June 2018 / Published online: 13 July 2018}

\maketitle

\begin{abstract}
This paper investigates the application of mesh adaptation techniques in the Non\hyp{}Ideal Compressible Fluid Dynamic (NICFD) regime, a region near the vapor\hyp{}liquid saturation curve where the flow behavior significantly departs from the ideal gas model, as indicated by a value of the fundamental derivative of gasdynamics less than one.
A recent interpolation-free finite-volume adaptive scheme is exploited to modify the grid connectivity in a conservative way, and the governing equations for compressible inviscid flows are solved within the Arbitrary Lagrangian Eulerian framework by including special fictitious fluxes representing volume modifications due to mesh adaptation.
The absence of interpolation of the solution to the new grid prevents spurious oscillations that may make the solution of the flow field in the NICFD regime more difficult and less robust.
Non\hyp{}ideal gas effects are taken into account by adopting the polytropic Peng-Robinson thermodynamic model.
The numerical results focus on the problem of a piston moving in a tube filled with siloxane \mdqm, a simple configuration which can be the core of experimental research activities aiming at investigating the thermodynamic behavior of NICFD flows.
Several numerical tests involving different piston movements and initial states in 2D and 3D assess the capability of the proposed adaption technique to correctly capture compression and expansion waves, as well as the generation and propagation of shock waves, in the NICFD and in the non\hyp{}classical regime.

%
\end{abstract}

\section{Introduction}
\label{intro}
Computational Fluid Dynamics (CFD) has undoubtedly become an important prediction, analysis, and design tool in many engineering applications.
New methods are continuously developed to better exploit computational resources, to increase solution accuracy, and to extend the range of applicability of CFD to diverse fields.
For instance, the numerical simulation of fluid flows 
within the so-called Non-Ideal Compressible Fluid Dynamics (NICFD) regime still represents some challenges for CFD experts, and it is a very active area of research~\cite{Vitale2015,Gori2017,Pini2017,Ameli2017,Gori2017a,Head2017,Keep2017}.

NICFD deals with flows occurring within the thermodynamic region wherein the fluid thermodynamic behavior significantly departs from the one predicted by the ideal gas model, as occurs for dense vapors, supercritical flows, and compressible two\hyp{}phase flows.
In this region, the attractive and repulsive molecular forces are not negligible and non\hyp{}ideal gas effects, such as non\hyp{}monotone variations of the Mach number along isentropic expansions, or non\hyp{}classical phenomena, such as rarefaction shocks, may occur~\cite{Colonna2006,Harinck2009,Nannan2014}.

Reliable simulations of NICFD flows require accurate thermodynamic models that are able to include non-ideal effects, but that are usually characterized by complex mathematical descriptions.
The inclusion of such thermodynamic models in CFD software is not a trivial task and leads to a considerable increase of the computational time, especially when the flow states are in the neighborhood of the Vapor-Liquid Equilibrium (VLE) curve~\cite{ColonnaRebay04}.

In recent years, different efforts have been made to extend standard techniques usually adopted under the perfect, i.e., polytropic ideal, gas assumption to the NICFD regime.
As a result, some CFD toolkits presently offer the capability to perform simulations of non-ideal compressible flows, such as for instance the open\hyp{}source multi\hyp{}physics software suite {SU2}, which has been recently equipped with a varied thermodynamic library for pure fluids~\cite{Vitale2015}, and commercial software such as {ANSYS CFD} or {STAR-CCM+}.

In addition, new methods are continuously developed to tackle specific numerical tasks or applications. For example, several contributions have been dedicated to numerical schemes~\cite{Vinokur1990,Abgrall1991,Arabi2017}, to efficient evaluation of thermodynamic quantities~\cite{Colonna2003,Cinnella2011}, and, more recently, to
shock waves~\cite{Pantano2017,Gori2017,Passmann2017}, and to turbulent flows~\cite{From2017,Sciacovelli2017} in the NICFD regime.
However, some numerical techniques widely used in ideal gas simulations have not been the subjects of an adequate investigation and assessment in the NICFD regime yet. In particular, in our opinion, an example of these shortcomings concerns mesh adaptation techniques in unsteady simulations.

Mesh adaptation is a valuable strategy for the simulation of flow fields characterized by different spatial scales and for unsteady simulations, where the position of the relevant flow features changes in space and/or time or the computational domain undergoes large deformations.
In these situations, the grid spacing can be locally reduced or increased to efficiently optimize solution accuracy, while preventing an excessive growth of the number of grid nodes.
Concerning the criteria used to modify the grid, different choices can be made, as for instance integral error indicators~\cite{Dwight2008,Fidkowski2011}, local a-posteriori error estimators based on interpolation error analysis~\cite{Formaggia2003,Coupez2011} or local indicators based on the behavior of the flow solution~\cite{Kallinderis1989,Choi2004,Kallinderis2017}.
Similarly, various adaptation strategies have been proposed, such as node displacement at fixed connectivity (r-adaptation)~\cite{Huang2011}, local grid-connectivity modification (r\hyp{}refinement)~\cite{Freitag1997,Mavriplis2000,Dapogny2014}, partial or complete re\hyp{}meshing, or modification of the polynomial order used to discretize the equations on each element (p-refinement)~\cite{Babuska1994,Dolejsi2015}, which can also be combined together, as hp-refinement.
However, standard mesh adaptation techniques require the interpolation of the solution from the old to the adapted grid.
As is well known, from a numerical point of view, solution interpolation in compressible flow is an undesirable operation, because it may undermine the accuracy of the solution and the properties of the scheme, such as stability and positivity-preserving properties, and it may introduce spurious oscillations. This latter fact can be particularly dangerous in NICFD simulations, where oscillations may bring the thermodynamic state of the fluid under the VLE curve, where the adopted single\hyp{}phase Equation of State (EoS) is invalid.

To overcome the difficulties related to the solution interpolation, the inviscid flow solver {Flowmesh} is currently under-development at the Department of Aerospace Science and Technology of Politecnico di Milano~\cite{flowmeshWeb}.
The solver implements a finite\hyp{}volume scheme for unstructured grids based on an innovative interpretation of  the modification of grid elements due to mesh adaptation within the Arbitrary Lagrangian\hyp{}Eulerian (ALE) framework \cite{Guardone2011,Isola2015,Re2017jcp}. Thanks to a series of fictitious continuous deformations of the finite volumes, interpolation of the solution from the initial to the adapted grid is avoided, thus preserving the properties of the fixed-connectivity ALE scheme while guaranteeing conservativeness by construction.
In our opinion, the absence in the proposed method of an explicit interpolation step has a favorable impact on NICFD simulations, where oscillations in the thermodynamic fluid state may undermine the robustness of the method to the point of jeopardizing the computation of the solution.
In addition, the flow solver is linked to the {VThermo} library, now included in {FluidProp}~\cite{FluidProp}, which offers the possibility of evaluating thermodynamic quantities by means of different EoSs, such as the van der Waals EoS~\cite{VanderWaals1873}, the Peng-Robinson EoS~\cite{PengRobinson1976}, the Martin-Hou EoS~\cite{MartinHou1955}, the Redlich-Kwong EoS~\cite{Redlich1949}, and the Soave--Redlich\hyp{}Kwong EoS~\cite{Soave1972}.
For what concerns mesh adaptation strategies, node displacement, insertion, deletion, and edge swap are exploited to locally modify the grid spacing, with the support of the automatic triangular and tetrahedral re\hyp{}meshers of the open-source library {Mmg}~\cite{mmgWeb}.

A natural application of a numerical tool with the outlined features is, in our opinion, the simulation of a flow inside a tube closed by movable pistons.
Different types of piston motions can be enforced to generate various flow fields, which may include shock waves, contact discontinuities, and smooth regions.
Indeed, this setup is often the core of experimental test\hyp{}rig arrangements thanks to its feasibility and flexibility~\cite{Bryson1965,Stalker1966,Kewley1974,Hannemann2016}.
However, the numerical simulation of these flow fields is challenging due to the presence of traveling waves with different spatial scales, and it represents a valuable test to assess the capabilities of the proposed adaptive method.

In essence, this paper describes the assessment of an adaptive Euler solver over 2D and 3D unstructured grids for moving body problems in the NICFD regime.
To authors' knowledge, this is the first application of unsteady mesh adaptation strategies to NICFD problems, and the interpolation\hyp{}free approach here exploited represents an added value with respect to standard adaptation techniques in this peculiar thermodynamic regime.
This work draws on the investigation of mesh adaptation criteria based on flow variables in the NICFD regime for bi-dimensional steady simulations described in~\cite{Re2017jet}.

The paper is organized as follows. Section~\ref{sec:TMD} outlines the main features of the thermodynamic modeling in the NICFD regime and the polytropic Peng\hyp{}Robinson model used in the simulation of the piston problems.
Section~\ref{sec:NumMeth} sketches the numerical method to solve the unsteady Euler equations over adaptive grids within the ALE framework.
Then, the adopted mesh adaptation techniques are described in Sec.~\ref{sec:adapt}, along with a summary of the entire computational process.
An assessment of the proposed unsteady method in 2D and 3D steady tests is given in Sec.~\ref{sec:assessment}.
Section~\ref{sec:results} describes the results of different piston problems.
More specifically, Subsec.~\ref{ssec:infinite2D} and \ref{ssec:finite} illustrate an oscillating piston in an infinite\hyp{}length and in a closed tube in the NICFD regime, starting from two different initial conditions. Then, we show the results for an impulsively started piston in a closed tube.
Subsection~\ref{ssec:nonClass} displays the results of two tests about the piston experiencing harmonic motion in the non\hyp{}classical region.
Three-dimensional results are presented in Subsec.~\ref{ssec:results3d}.
Finally, the conclusions of the present work are drawn in Sec.~\ref{sec:concl}.

\section{Thermodynamic modeling in NICFD regime}
\label{sec:TMD}
In NICFD, a quantitative measure of the deviation of the flow behavior from the ideal gas model is given by the value of the fundamental derivative of gasdynamics $\Gamma$~\cite{Thompson1971},
\begin{equation}
\label{e:Gamma}
\Gamma = 1 + \dfrac{\rho}{c} \left(\dfrac{\partial c}{\partial \rho} \right)_s \,,
\end{equation}
which expresses the isentropic variation of the sound speed, being $\rho$, $c$, $s$ the density, the sound speed and the specific entropy, respectively.

The occurrence of non-ideal gas effects is determined by values of $\Gamma$ less than unity.
More specifically, the gasdynamic behavior of fluids characterized by $0<\Gamma <1$ shows qualitative differences with respect to the ideal one. The most peculiar is the non-monotone dependence of the Mach number on the density along isentropic expansions because of the increase of the speed of sound~\cite{Cramer1991,Kluwick1993}.
For thermodynamic states at $\Gamma<0$, non-classical gasdynamic phenomena, such as rarefaction shock waves and composite waves, may occur~\cite{Cramer1984,Menikoff1989},
although no experimental evidence of the existence of such exotic flow features in single\hyp{}phase vapor flows is available.
The value of the fundamental derivative varies over the thermodynamic plane and some fluids have a finite region near the VLE curve where $\Gamma<1$~\cite{Harinck2009}.
However, recent advancements have predicted the possible occurrence of even non-classical phenomena near the critical point also for quite simple molecules~\cite{Nannan2014,Nannan2016}.

Given the inappropriateness for NICFD simulations of the perfect, or so\hyp{}called Polytropic Ideal Gas (PIG) model---which predicts a constant $\Gamma^{\mathsf{PIG}}=(\gamma+1)/2>1$, where $\gamma=c_p/c_v>1$ is the constant ratio between the isobaric ($c_p$) and isochoric ($c_v$) specific heat capacities---a non\hyp{}ideal thermodynamic model is required to describe the flow behavior.
For pure fluids, the complete evaluation of the thermodynamic state can be obtained from a thermodynamic potential expressing directly the fundamental relation, as the Helmholtz free energy, or from two compatible equations of states~\cite{GuardoneArgrow2005}.

Different modeling approaches have been proposed to describe the thermodynamic behavior of the flow in CFD simulations.
For instance, the pressure $P$ can be expressed in terms of temperature $T$ and specific volume $v$ by a cubic equation, generally composed by two terms, one accounting for the attractive intermolecular forces and one for the repulsive covolume effects.
The simplest cubic EoS is the van der Waals EoS, but several different pressure EoSs have been proposed~\cite{PengRobinson1976,MartinHou1955,Redlich1949,Soave1972}.
To have a complete thermodynamic description, the pressure EoS has to be complemented by a compatible caloric EoS~\cite{Callen1985}.
A different approach may rely on an accurate multi\hyp{}parameter EoS expressing the fundamental relation, see for instance~\cite{Span2003,Lemmon2006}.
Because of their high non-linearity and complexity, a direct implementation in CFD codes of such EoSs is practically unaffordable, but they can be used in a preceding step to build accurate Look-Up Tables which are then recalled during the simulation to evaluate the thermodynamic state by interpolation~\cite{Rinaldi2014,Pini2015,Moraga2017}.

\subsection{Peng-Robinson equation of state model}
\label{ssec:PReos}
In this work, to model the thermodynamic behavior of the flow we use the pressure
Peng\hyp{}Robinson EoS~\cite{PengRobinson1976}:
\begin{equation}
\label{e:pengRob}
P(T,v) = \dfrac{R T}{v -b} - \dfrac{a \, \alpha_\omega(T_\mathrm{r})^2}{v(v+b) + b(v-b)}\, ,
\end{equation} 
where $a$ and $b$ are two fluid-specific constants given in terms of the critical pressure $P\crit$ and temperature $T\crit$,
\[ a =0.45724 \, \dfrac{R^2 T\crit^2}{P\crit} \, , \qquad 
   b =0.0778  \, \dfrac{R T\crit}{P\crit}  \, , \]
and $\alpha_\omega$ is a dimensionless function of reduced temperature $T_\mathrm{r}=T/T\crit$, namely
\[ \alpha_\omega(T_\mathrm{r}) = 1 + f_\omega \left(1-\sqrt{T_\mathrm{r}} \right) \, \]
where $f_\omega$ is a constant characteristic of the substance, defined in terms of the acentric factor $\omega$,
\[ f_\omega = 0.37464 + 1.54226\, \omega - 0.26699 \, \omega^2 \,.\]
The acentric factor is a parameter that takes into account the polarity and the lack of spherical shapes of the molecules. 
In this work, its (approximate) value is computed by means of the Edmister's equation~\cite{Poling2001}:
\begin{equation}
\label{e:acentricFactor}
\omega = \dfrac{3}{7} \left( \dfrac{T_\mathrm{b}}{T\crit - T_\mathrm{b}} \right) \; \log_{10} 
  \left(\dfrac{P\crit}{101325} \right) \, ,
\end{equation}
where $T_\mathrm{b}$ is the temperature at the normal boiling point.
 
Given the pressure EoS, the compatible caloric EoS for the internal energy per unit of mass $e=e(T,v)$ can be derived by integrating the reciprocity relation 
\begin{equation}
\label{e:reciprocity}
\left( \dfrac{\partial e(T,v)}{\partial v}\right) =
T \left(\dfrac{\partial P(T,v)}{\partial T}\right)_v - P(t,v) \,.
\end{equation}
However, the integration with respect to the specific volume allows to determine the compatible EoS only up to an integration constant function of temperature, which is usually connected to the dependence of the energy on temperature in the dilute gas limit.
In common practice, it is defined as
\begin{equation}
\label{e:phiT}
\phi(T) =\int_{T_{\mathrm{ref}}}^T c_{v_\infty}(\tau) \, \mathrm{d} \tau \, ,
\end{equation}
where $c_{v_\infty}$ is the isochoric specific heat capacity in the ideal gas limit, that is for $v \to \infty$. For more details, see Ref.~\cite{GuardoneArgrow2005}.

The caloric EoS compatible to Eq.~\eqref{e:pengRob} is
\begin{multline}
e(T,v) = e_{\mathrm{ref}} + \phi(T) - \dfrac{a}{b}\dfrac{\alpha_\omega(T_\mathrm{r})}{2\sqrt{2}} \, \Bigl[\alpha_\omega(T_\mathrm{r}) \\
+ f_\omega \sqrt{T_\mathrm{r}} \, \Bigr] \, \ln \dfrac{1+\sqrt{2} + v/b}{1-\sqrt{2} + v/b}
\end{multline}
where $e_{\mathrm{ref}}$ is an arbitrary reference value for the specific energy, used in the integration of Eq.~\eqref{e:reciprocity}.
In this work, we assume a polytropic behavior in the dilute gas limit ($v \to \infty$), therefore $\phi(T)$ is a linear function of the temperature, i.e., $c_{v_{\infty}}$ is constant.

A convenient parameter that easily expresses the deviation of the pressure EoS from the ideal gas model is the compressibility factor $Z=Pv/(RT)$, where $R$ is the gas constant. For the PIG model, $Z$ is constant and equal to 1. 
The Peng\hyp{}Robinson EoS predicts the same compressibility factor at the critical point for all substances. Indeed, by evaluating Eq.~\eqref{e:pengRob} at $P\crit=P(T\crit, v\crit)$, a cubic equation for $Z\crit$ is obtained, which has only one real root $Z\crit^{\mathsf{PR}}=0.3214$.
As a consequence, this EoS, as all the cubic ones, is not able to correctly compute all three critical point coordinates $P\crit$, $T\crit$, and $v\crit$~\cite{GuardoneArgrow2005}.

\section{Numerical method}
\label{sec:NumMeth}
The governing equations for unsteady inviscid compressible flows are provided by the Euler equations.
For a control volume $\CC$ within the spatial domain $\Omega$ they read
\begin{equation}
\label{e:euler}
 \dfrac{\mathrm{d}}{\mathrm{d}t} \int_{\CC} \! \uu \; \mathrm{d}\xx +
   \oint_{\dCC}  \!\!\!  \ff(\uu)  \cdot \nn \; d\mathbf{s} = 0 \, , 
\end{equation}
where $\xx \in \mathbb{R}^k $ and $t>0$ are as usual the position and the time, $\uu = [ \rho, \, \mm, \, \Et  ]^{\mathrm{T}}$ is the vector of conservative variables (density, momentum density, and total energy density, respectively) and $\nn(\xx,t)$ denotes the outward unit vector normal to the boundary $\dCC$, over which the position vector is denoted $\mathbf{s} \in \mathbb{R}^{k-1} $.
The inviscid flux function $\ff(\uu)$ is defined as
\[ 
   \ff(\uu) = \left[
 \mm , \;
 \mm  \otimes \mm / \rho + \Pi(\uu) \mathbb{I}^k  ,\;
 \left[ \Et +  \Pi(\uu) \right] \mm/\rho 
 \right]^{\mathrm{T}} \, ,
\]
where $\Pi(\uu)$ is the pressure function and $\mathbb{I}^k$ is the $k \times k$ identity matrix.
Suitable initial and boundary conditions have to be specified to complement Eq.~\eqref{e:euler}~\cite{LeVeque1992}.

The thermodynamic model completes the Euler equations through the pressure function $\Pi$, which expresses the pressure $P$ as a function of the conservative variables. Indeed, the pressure and caloric EoSs are manipulated so that
\[P=P(e(T,v),v) = P(E, \rho) = P\left(\Et - \frac{\Vert \mm \Vert^2} {2\, \rho} , \rho\right) = \Pi(\uu) \, ,\]
where $E$ is the internal energy density (per unit of volume).

\subsection{Spatial and temporal discretization}
\label{ssec:discretization}
The discrete form of the Euler equations is now briefly described. Since the focus here is on the influence of the thermodynamic model on the method, only the relevant results of the discretization process are illustrated. A detailed description of the scheme can be found in~\cite{Guardone2011,Isola2015} for the 2D and in~\cite{Re2017jcp,RePhd2016} for the 3D cases.

A node\hyp{}centered edge\hyp{}based finite\hyp{}volume scheme is exploited to spatially discretize the governing equations over the computational grid. Thus, the domain is split in a finite number of non-overlapping finite volumes $\Ci$ so that $\bigcup_i \Ci(t)= \Omega(t)$, where $\Ci$ is a volume surrounding the grid node $i$.
By adopting the backward Euler scheme for time integration, the discretized governing equations can be written as
\begin{equation}
\label{e:eulerDisc}
\dfrac{\ui\tnn \!\!- \ui\tn} {\Delta t} \, V_i = 
\!\! \sum_{k \in \KK_{i,\neq}} \!\! \Phi(\ui, \uu_k, \etab_{ik})\tnn + 
\Phi^\partial(\ui, \xib_i )\tnn \,,
\end{equation}
where $V_i$ is the volume of the cell $\Ci$, $\uu_i$ is the average solution $\uu$ over $\Ci$, $\KK_{i,\neq}$ denotes the set of finite volumes different from $\Ci$ that share a portion of their boundary with $\Ci$; $\Phi$ and $\Phi^\partial$ are suitable integrated numerical fluxes that represent, respectively, the flux across the domain cell interface $\dCC_{ik}=\dCC_i \cap \dCC_k$ and, if the node $i$ lies on the boundary, across the boundary interface $\dCC_{i,\partial}=\dCC_i \cap \partial\Omega$. 
The superscripts $n$ and $n+1$ indicate variables evaluated at time $t\tn$ and $t\tnn$, respectively, while $\Delta t$ is the time step in-between.
Moreover, we have introduced the integrated normal vectors:
\begin{equation}
\etab_{ik}=\int_{\dCC_{ik}}  \!\!\!\!\!\!  \nn_i \; \mathrm{d} \xx
\qquad \mathrm{and} \qquad
\xib_i=\int_{\dCC_{i,\partial}}  \!\!\!\!\!\!\!\!\!  \nn_i \; \mathrm{d} \mathbf{s} 
\end{equation}
with $\nn_i$ denoting the outward normal with respect to the finite volume $\Ci$.

A pseudo\hyp{}time step method is used to solve the non\hyp{}linear Eq.~\eqref{e:eulerDisc} and the solution at time $t\tnn$ is obtained through an iterative process.
The pseudo\hyp{}time derivative is discretized by means of the backward Euler scheme too, and the unsteady residual,  containing the numerical fluxes at the actual pseudo\hyp{}time step, is approximated by a Taylor expansion. Moreover, according to the defect\hyp{}correction approach, the exact Jacobian of the integrated fluxes is approximated by the Jacobian of the first order fluxes only, to increase its diagonal dominance~\cite{Koren1988}.
At each pseudo-time step, a Symmetric Gauss Seidel method is used to solve the system of linear equations.
Further details about the time integration can be found in~\cite{IsolaPhd,Carpentieri2009}.

\subsubsection{Integrated numerical fluxes}
For the integrated flux $\Phi$, a high\hyp{}resolution scheme based on the Total Variation Diminishing (TVD) approach~\cite{Harten1997} is used in the present work to correctly capture shock waves and contact discontinuities within the NICFD regime of interest.
More specifically, near discontinuities the second\hyp{}order centered approximation of the interface flux is replaced by the first order monotonicity\hyp{}preserving Roe scheme. The switch is controlled by the van Leer limiter.
By denoting the limiter $\Psi= \mathrm{diag}\left\lbrace\Psi_1 \dotsc \Psi_{k+2}\right\rbrace$, the integrated numerical flux across the interface $\dCC_{ik}$ reads
\[ \Phi = \dfrac{\ff(\uu_i) + \ff(\uu_k)}{2} \cdot \etab_{ik}
 + \dfrac{1}{2}\mathsf{R}\vert \widetilde{\mathsf{\Lambda}} \vert \left( \Psi - \mathbb{I}\right) \mathsf{L}(\uu_k -\uu_i)
\]
where $\widetilde{\mathsf{\Lambda}}$ is the Roe matrix, defined here as the Jacobian of the flux function $\ff(\uu)$ projected along the normal direction $\hat{\etab}_{ik}$ and evaluated at the intermediate Roe state $\tilde{\uu}=\tilde{\uu}(\uu_i, \uu_k)$.

Under the polytropic ideal gas assumption, only two variables are required to define the Roe matrix~\cite{Roe1981}. Conversely, if a non\hyp{}ideal thermodynamic model is assumed, the Roe matrix is not uniquely determined by the conservation property~\cite{Guardone2002} and several approaches were proposed to compute the Roe matrix. Most standard approaches rely on an augmented intermediate state, which includes the pressure derivatives as additional variables and aims at obtaining a quasi-Jacobian form of the Roe matrix~\cite{Glaister1988,Vinokur1990,Abgrall1991,Cox1994}.
However, the average thermodynamic derivatives do not retain their proper significance and this may lead to inconsistencies whenever they are used to compute other thermodynamic quantities, such as the speed of sound~\cite{Toumi1992,Guardone2002}.
A different approach is proposed in~\cite{Guardone2002}, where the Jacobian form of the Roe matrix is enforced. This choice results in an intermediate state that is a one-parameter family of solutions, and the following additional condition is imposed to determine the intermediate density:
\[\left(\dfrac{\partial P}{\partial E}\right)_{\!\!\!\! \rho} \bigl(E_i - E_k \bigr) + 
 \left(\dfrac{\partial P}{\partial \rho}\right)_{\!\!\!\! E} \bigl(\rho_i - \rho_k \bigr) = P_i - P_k \, .\]
Unfortunately, when using complex EoSs the computation of the intermediate density from the previous relation is not easy and it requires numerical techniques~\cite{Guardone2007a}.

Nevertheless, it has been shown that no relevant differences can be appreciated in the results when using different, simpler approaches~\cite{Mottura1997,Guardone2007a}. Therefore, the generalized Roe matrix for non-ideal gas flows is computed here following the simplified approach proposed by Cinnella~\cite{Cinnella2006}, which consists in selecting the intermediate density as $\widetilde{\rho}=\sqrt{\rho_i \, \rho_k}$, and the intermediate velocity and total enthalpy through the same definitions used for the PIG model, which are valid also for non-ideal gases as proved in~\cite{Guardone2002}.

\subsubsection{Boundary fluxes}
The boundary conditions are imposed in a weak form, i.e., by evaluating the flux in a suitable boundary state $\uu^\partial_i= \uu^\partial(\uu_i, \mathbf{b})$, function of the solution on the boundary node $\uu_i$ and of the boundary data $\mathbf{b}$.

In the present work, three types of boundary conditions are imposed: the slip wall, the normal inlet/outlet, and the non\hyp{}reflecting free surface.
The former one is imposed by setting the component of the fluid velocity normal to the boundary equal to the boundary velocity along the normal direction.
The second condition models an infinite-length open-end tube, that does not influence the flow inside the domain but preserves only the flow direction parallel to the tube axis, without imposing the sense (inflow or outflow). Hence, this boundary condition simply nullifies the component of the fluid velocity tangential to the open-boundary (or, if need be, sets it equal to the tangential velocity of the boundary).
For the latter case, the boundary state is computed via characteristic reconstruction, and the correct number of the physical variables that can be imposed is automatically computed by the eigenvalues analysis~\cite{Selmin1993,Vitale2015}.

Because an implicit pseudo\hyp{}time method is adopted in this work and a Taylor expansion is used to approximate the residual terms containing the numerical fluxes, the evaluation of the Jacobian matrix at the boundary state and the derivative $\frac{\partial \uu^\partial_i}{\partial \uu_i}$ are also required~\cite{IsolaPhd}.

\subsection{ALE framework for dynamic grids}
Unsteady problems often require updating the computational domain to follow the boundary motion or deformation.
In such situations, it is common practice to redistribute the displacement occurring at the boundary among the internal grid nodes depending on their shape and volume without modifying the grid connectivity, see for instance~\cite{Batina1990,Venkatakrishnan1996,Degand2002}.
The same technique can be used to move grid nodes to better describe the local flow features, namely for r-adaptation.
As a consequence, on a dynamic grid, the formulation of the flow equations has to be modified to account for the relative motion of the grid with respect to the fluid.
The Arbitrary Lagrangian-Eulerian (ALE) formulation is a widely-used and effective strategy to accomplish this task~\cite{Hirt1974,Donea1982}.
Within the ALE framework, the governing equations are enforced over control volumes that can move and deform independently of the fluid velocity. 

The ALE formulation of the Euler equations reads
\begin{equation}
\label{e:eulerAle}
 \dfrac{\mathrm{d}}{\mathrm{d}t} \int_{\CC(t)} \!\!\!\!\! \uu \; \mathrm{d}\xx +
   \oint_{\dCC(t)}  \!\!\!\!\!\!\!  \left[ \ff(\uu) - \uu \vv \right] \cdot \nn \; \mathrm{d}\mathbf{s} = 0 \, , 
\end{equation}
where $\vv$ is the velocity of the control volume.
The previous equation can be discretized through the same strategy outlined in Sect.~\ref{ssec:discretization}, provided that the contribution of the grid velocity is taken into account in the integrated fluxes. In this regard, two additional quantities labeled as interface velocities are defined as
\begin{equation}
\label{e:interfaceVel}
\nu_{ik}=\int_{\dCC_{ik}}  \!\!\!\!\!\!  \vv \cdot \nn_i \; \mathrm{d} \xx
\qquad \mathrm{and} \qquad
\nu_i=\int_{\dCC_{i,\partial}}  \!\!\!\!\!\!\!\!\!  \vv \cdot \nn_i \; \mathrm{d} \mathbf{s} \, ,
\end{equation}
and the ALE formulation of Eq.~\eqref{e:eulerDisc} can be written as
\begin{multline}
\left({V_i\tnn \ui\tnn - V_i\tn \ui\tn}\right) \big/{\Delta t} = \\
\sum_{k \in \KK_{i,\neq}} \!\! \phi(\ui, \uu_k, \etab_{ik}, \nu_{ik} )\tnn + 
\phi^\partial(\ui, \xib_i, \nu_{i} )\tnn \,.
\end{multline}

The additional constraint known as the Geometrical Conservation Law (GCL) is beneficial to avoid spurious oscillations and instabilities~\cite{Formaggia2004,Mavriplis2006,Etienne2009}.
The GCL states that the movement of the computational grid should not affect a uniform solution and it amounts to a suitable computation of the geometrical quantities involved in the grid movement~\cite{Lesoinne1996}.
In this work, the interface velocities defined in Eq.~\eqref{e:interfaceVel} are computed in a GCL-compliant fashion by means of the following relations:
\begin{equation}
\label{e:dV}
\Delta V_{ik}\tnn = \Delta t \, \nu_{ik}\tnn
\qquad \mathrm{and} \qquad
\Delta V_{i,\partial}\tnn = \Delta t  \, \nu_i\tnn \, ,
\end{equation}
where $\Delta V_{ik}\tnn$  and $\Delta V_{i,\partial}\tnn$ are the volumes swept during the time step $\Delta t$ by the interfaces $\dCC_{ik}$ and $\dCC_{i,\partial}$, respectively.
Assuming that the positions of the grid nodes at the beginning and at the end of the time step are given, the swept volumes can be easily computed by exploiting geometrical relations, as shown in~\cite{Guardone2011} for triangular grids and in~\cite{Re2017jcp} for tetrahedral ones.

\subsubsection{Extension to variable connectivity grids}
\label{ssec:threeSteps}
\begin{figure*}\sidecaption
\includegraphics*{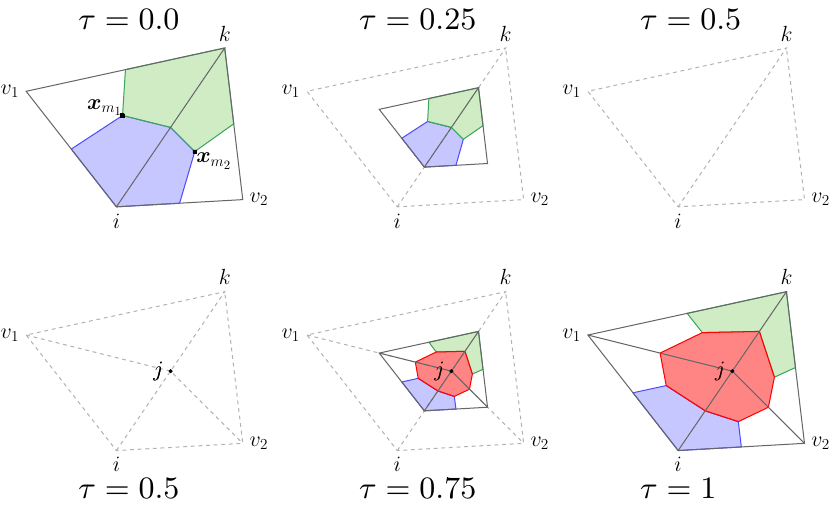}
\caption{
Three-step procedure applied to the split of edge $e_{ik}$. The non-dimensional time $0 \leq \tau \leq 1$ is used to describe the different fictitious steps.
The dashed grey lines show the grid connectivity in the original/final configuration, while the portions of the finite volumes associated to $i$, $k$, and $j$ are shown with light green \square{draw=DarkGreen,fill=LightGreen!60},
light blue \square{draw=MediumBlue,fill=LightSteelBlue},
and orange \square{draw=red,fill=red!50}, respectively.
The label $\xx_{m_i}$ indicates the barycenter of the element $m_i$.
In the first row the collapse phase $0< \tau < 0.5$ is depicted: the elements that share the edge $e_{ik}$ at $\tau=0$ are collapsed over its mid-point. When they reach a null area, the connectivity is changed ($\tau=0.5$):
the new point $j$ is inserted, the edge $e_{ik}$ is split into two edges ($i$-$j$ and $k$-$j$), and two new edges are created to connect $j$ to $v_1$ and $v_2$.
The second row displays the expansion phase $0.5< \tau < 1$: the nodes $i$, $k$, $v_1$, $v_2$ return to their original positions to reach the final configuration (at $\zeta=1$).
}
\label{f:threeSteps}
\end{figure*}

The ALE scheme presented in the previous subsection has been extended also to variable connectivity grids.
Previous works~\cite{Guardone2011,Isola2015,Re2017jcp} detail how to compute the swept volumes $\Delta V_{ik}\tnn$ and $\Delta V_{i,\partial}\tnn$ when the local connectivity varies due to mesh adaptation.
Connectivity changes are described as fictitious continuous deformations of the involved finite volumes  through a series of collapse and expansion operations.
More precisely, the following three steps can be defined with the aid of the fictitious dimensionless time~$\tau$ (see also Fig.~\ref{f:threeSteps} for an example about edge split in 2D).
\begin{enumerate}
\item \textit{Collapse}, $0 \!<\! \tau \!<\! 0.5$: all elements involved in the local adaptation collapse into an arbitrary point.
\item \textit{Connectivity change}, $\tau \!\!=\! 0.5$: when all involved elements reach null volumes, nodes may be inserted or deleted.
\item \textit{Expansion}, $0.5 \!<\! \tau \!<\! 1$: all elements still active (i.e., not deleted at $\tau=0.5$) expand to the final configuration.
\end{enumerate}
The collapse and expansion phases are merely continuous deformations from the initial position to the pivotal point, and from it to the final position. Since the positions of these points are known, the volume swept by the involved interfaces can be easily computed from geometrical considerations. Moreover, the fictitious fluxes generated by these deformations can be expressed within the ALE framework as shown for standard fixed-connectivity deformation and included in Eq.~\eqref{e:eulerAle}.
Conversely, the connectivity change at $\tau\!\!=\! 0.5$ occurs while the elements have null volume, so no volume is swept by any interface and no flux is exchanged.
In this way, we are able to compute the GCL-compliant interface velocities through Eqs.~\eqref{e:dV} also when local grid topology is modified and to include the volume changes due to mesh adaptation into the standard ALE scheme, without undermining the fixed-connectivity properties and enforcing conservativeness.

The three steps and the core of the procedure are the same for node insertion, node deletion, or edge swap, both in 2D and 3D; only the involved elements depend on each specific grid modification.
The three-step procedure is carried out immediately after each connectivity modification and it is instantaneous, i.e., it starts and ends during the same time step $\Delta t$ at which the grid is adapted.
Finally, we remark on the local character of the three\hyp{}step procedure. The interfaces that are not modified by the connectivity change do not take part in three\hyp{}step procedure, as their contribution would be null since they
would sweep the same volume but with different sign during the collapse and the expansion phase~\cite{Re2017jcp}.

\section{Mesh adaptation in unsteady NICFD simulations}
\label{sec:adapt}
The accuracy of the numerical solution of the flow equations is strongly influenced by the grid spacing.
For simple geometries or in steady simulations, it is possible to generate a priori a mesh that provides sufficiently small numerical errors with respect to the modeling errors. 
This task becomes considerably more challenging or impossible for complex flow fields characterized by different spatial scales and in unsteady simulations, where the grid spacing has to be related to the behavior of the solution in an efficient way.
Solution\hyp{}dependent adaptive grid techniques represent a valuable tool to deal with this requirement.
More specifically, in unsteady simulations, mesh adaptation techniques may be profitably used to tackle two different tasks:
to handle large deformations of the boundaries, and to capture the relevant flow features that originate and move through the domain~\cite{Johnson1999,Hassan2000,Venkatakrishnan1996}, as outlined in the following subsections.

\subsection{Adaptation on flow features}
\label{ssec:solAdapt}
The solution accuracy is here optimized following the equi\hyp{}distribution principle, whereby the grid spacing is modified to equi\hyp{}distribute the error over
the mesh~\cite{Borouchaki1997,Dolejsi1998}.
The first step of this strategy consists in the definition of an error estimator, based upon the flow solution.
It appears evident that an effective error indicator will be extremely advantageous to the entire adaptation process.
In the NICFD regime, the error estimators based on the Mach number seem to be more efficient than the density\hyp{}based ones because of the non-monotone dependence of the Mach number on density along isentropic expansions for supersonic flows at $\Gamma<1$~\cite{Re2017jet}.

A metric map is used to prescribe the size of grid elements during the mesh modification process.
If anisotropic mesh adaptation is performed, the shape and the orientation of the elements are also prescribed~\cite{Borouchaki1997,DelPino2011,Coupez2011}.
Two different procedures are followed to build isotropic or anisotropic metrics. In the former case, a target grid spacing is computed at each grid node by reducing (or augmenting) the actual average size of the edges connected to the node if the error is greater (or less) than a refinement (coarsening) threshold, prescribed in terms of mean and standard deviation of the estimated error over the domain.
Conversely, the anisotropic map is built on the basis of the eigenstructure of the Hessian matrix of a certain solution variable~\cite{FreyAlauzet2005}, as 
$\mathcal{M} = {R} \widetilde{\Lambda} L ,$ where $R$ and $L$ are the right and left eigenvectors and
\[ \widetilde{\Lambda} =\mathsf{diag} \left\lbrace \min \left( \max \left( \dfrac{c \lvert \lambda_p \rvert}{\epsilon}, \dfrac{1}{\ell_{\mathrm{max}}^2} \right), \dfrac{1}{\ell_{\mathrm{min}}^2} \right)\right\rbrace \,,\]
with $\lambda_p$ the $p$-th eigenvalue,
$\ell_{\mathrm{max}}$/$\ell_{\mathrm{min}}$ the maximum/ minimum edge length,
$c=\frac{k^2}{2(k+1)^2}$ a constant, and
$\epsilon$ a user\hyp{}defined threshold for the maximum acceptable error, which in all presented computations ranges from $10^{-7}$ to $10^{-6}$.
More details about the construction of the metric maps can be found in~\cite{RePhd2016,Re2017aiaa,Re2017coupled}.

The metric $\mathcal{M}(\xx)$ is a field of symmetric positive matrices $\mathbb{R} ^{k\times k}$ that defines a Riemannian structure over $\Omega$.
The length of a vector $\mathbf{w}$ in terms of this map is given by 
\begin{equation}
\label{e:lengthM}
\lVert \mathbf{w} \lVert_{\mathcal{M}} =
  \sqrt{  \mathbf{w}{^\mathrm{T}}  \mathcal{M} \mathbf{w} \,} \,.
\end{equation}
and the goal of mesh adaptation is to obtain a unit mesh with respect to this metric, i.e., a mesh such that
$\lVert \mathbf{w} \lVert_{\mathcal{M}} = 1$ for all edges.

Given the initial grid and the metric $\mathcal{M}$, the open-source library \mmg~\cite{mmgWeb,Mmg2008} performs automatically a series of different local modifications to make the initial grid as similar as possible to a unit mesh~\cite{Dapogny2014}.
At first, the length of all grid edges is computed according to Eq.~\eqref{e:lengthM}, then the grid is locally modified both on the interior and on the boundary.
Where the edges are too long, the grid spacing is reduced by inserting a new node via element or edge split, or Delaunay triangulation. On the contrary, the nodes connected to edges having $ \lVert \mathbf{w} \lVert_{\mathcal{M}} \gg 1 $ are removed via edge collapse.
Edge swapping and barycentric regularization are also used.
In addition, the gradation control technique~\cite{Borouchaki1998} is exploited to limit the variation in size among adjacent grid elements.
All these operations are performed only if they lead to an improvement of the quality of the involved elements, which is defined as
\begin{itemize}
\item for tetrahedra: $\;Q_m = \alpha \, V_m / \big( \, \textstyle{\sum}_{i=1}^6 \; \ell_i^2\, \big)^{3/2} $,
\item for triangles: $\;Q_m = \ell_{\max} \; \sum_{i=1}^3 \; \ell_i \big / (2\;V_m) $,
\end{itemize}
where $V_m$ is the volume (or area in 2D) of the element $m$, $\ell_i$ is the edge length, $\ell_{\max}$ is the longest edge in the triangle, and $\alpha$ is a constant introduced to give $Q_m=1$ for a regular tetrahedron.
It is not possible to specify a maximum or minimum number of grid nodes, but the effect of mesh adaptation can be controlled through the parameter $\epsilon$ in the anisotropic case, and through the refinement/coarsening thresholds in the isotropic case.

\subsection{Large boundary displacements}
\label{ssec:boundAdapt}
Mesh adaptation here is exploited also to cope with large boundary movements.
At each time step, after the boundary nodes are moved to the new positions, the internal nodes are redistributed keeping the grid connectivity fixed. 
According to the elastic analogy~\cite{Batina1990}, the largest grid elements account for the major part of the deformation while the smallest ones move almost rigidly.
However, when the displacements are large, this strategy may lead to badly-shaped or null-volume elements.
In such situations, mesh adaptation is exploited to restore grid quality. 

As detailed in~\cite{Re2017aiaa}, the displacement to be imposed at the boundary nodes during the time step (labeled as $\dxB$) is parameterized by a linear function.
If the elastic analogy fails in producing a valid mesh, the displacement is split into smaller portions and the largest portion of $\dxB$ that leads to a valid mesh is imposed. Then, mesh optimization is carried out aiming only at increasing the quality $Q_m$ of all elements, not at modifying the grid spacing.
In this specific application of mesh adaptation, the edge swapping technique is particularly effective to restore the element quality, while keeping the number of grid nodes constant~\cite{Freitag1997}.

\subsection{Summary of the computational procedure}
\label{ssec:summary}
\begin{figure}
  \includegraphics[width=\hsize]{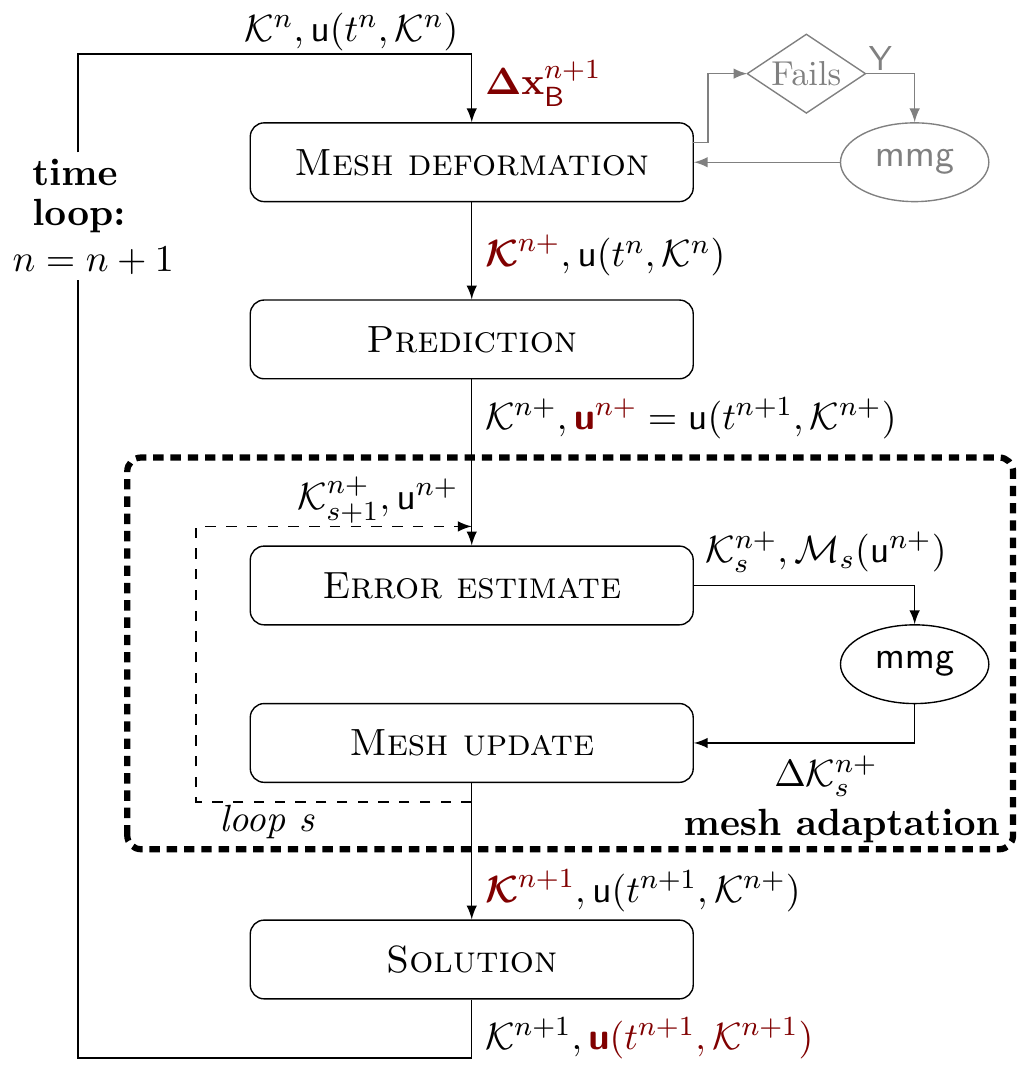}
\caption{Whole adaptive computational procedure for unsteady problems.
Mesh deformation is performed: if elastic analogy fails, \mmg\ is used to modify grid connectivity.
The grid $\KK^{n+}$ complies with the boundary displacement $\dxB$, and over it the solution $\uu^{n+}$ is computed.
The metric field $\mathcal{M}(\uu\tns)$ is passed as input to \mmg\ which communicates to the flow solver all performed modifications $\Delta \KK^{n+}$, so that it can compute the swept volumes $\Delta V$.
The mesh adaptation procedure, highlighted by the thick dashed line, can be repeated (\textit{loop s}).
Finally, the solution at $t\tnn$ over the adapted grid $\KK\tnn$ is computed.
The colored and bold symbols denote the variables changed at each step.}
\label{f:compProc}       
\end{figure}

To summarize, the main steps of the computational procedure adopted for unsteady adaptive simulations are outlined, with reference to Fig.~\ref{f:compProc}.
The generic time step $t\tn\leq t \leq t\tnn$ begins with the grid $\KK\tn$ and the corresponding solution $\uu(t\tn,\KK\tn)$.
Then, the following operations are performed.
\begin{description}
\item[\textit{Mesh deformation:}] the movement $\dxB$ is imposed at the boundary and the internal nodes are consequently relocated. If the elastic analogy fails, a mesh optimization stage including edge swapping is performed (see Sec.~\ref{ssec:boundAdapt}). The resulting grid is labeled $\KK\tns$ to highlight that it is intermediate, namely it complies with the new geometry but the solution is the previous one.
\item[\textit{Prediction:}] the solution $\uu\tns=\uu(t\tnn, \KK\tns)$ is computed over the new grid. This step prevents a delay between the solution\hyp{}based mesh adaptation and the actual geometry.
\item[\textit{Error estimate:}] the metric map $\mathcal{M}$ is generated on the basis of the solution $\uu\tns$ as summarized in Sec.~\ref{ssec:solAdapt}.
\item[\textit{Mesh adaptation:}] the grid $\KK\tns$ and the metric map are used as inputs to the library \mmg, which locally adapts the grid through node insertions, deletions, relocations, and edge swapping.
\item[\textit{ Mesh update:}] the flow solver receives from \mmg\ all performed local modifications $\Delta\KK\tns$,
that are needed to update the finite-volume discretization and to compute the swept volumes $\Delta V_{ik}$, $\Delta V_{i,\partial}$ by means of the three-step procedure (see Sec.~\ref{ssec:threeSteps}).
\item[\textit{Solution:}] the final solution $\uu\tnn$ on the grid $\KK\tnn$ is computed, using the solution $\uu\tns$ as initial guess.
\end{description}
Steps three to five can be repeated to perform multiple adaptation cycles, as indicated by the \emph{loop s} in Fig.~\ref{f:compProc}.

The use of the re\hyp{}mesher library \mmg~\cite{mmgWeb,Mmg2008} alleviates the inherent difficulties of mesh adaptation over unstructured three\hyp{}dimensional grids, which, for instance, requires to be able to deal with a variable number of elements involved in each local grid modification or with the multiple options that are available for an edge swapping.
The choice to exploit an external library to cope with these tasks allows to implement mesh adaptation techniques into the flow solver efficiently, although some modifications to the \mmg\ library have been required to exploit the three-step procedure of Sec.~\ref{ssec:threeSteps}.

\section{Assessment of the unsteady adaptive method for in NICFD regime}
\label{sec:assessment}
This section presents the first validation of the proposed unsteady method for 2D and 3D problems in the NICFD regime.
For the validation of the steady version, the reader is referred to~\cite{Re2017jet}.
The geometries of the test cases consist in a simple symmetric wedge in 2D and in a portion of a cone in 3D, exposed to a free stream.
As it is known, a steady oblique (conical, in 3D) shock wave is generated to deflect the flow such that it is aligned with the downstream wall.
This steady problem can be investigated by means of an unsteady approach if we consider the far\hyp{}field quiet and the wedge (cone, in 3D) moving at the free stream velocity,
namely, if we switch from the body to the laboratory reference frame.
Obviously, the thermodynamic variables that are independent from the reference frame, such as the pressure, are the same in the steady and unsteady simulation.

\subsection{Unsteady simulations of an oblique shock}
\label{ssec:oblique}
\begin{table}
\caption{Test conditions for the oblique shock test, and thermodynamic properties of the working fluid $\mathrm{MDM}$.
The deviation angle due by the wedge is $\theta=20^\circ$ and the oblique shock angle is $\beta=37.60^\circ$~\cite{Gori2017}.}
\label{t1:mdm} 
\begin{tabular}{llll}
\hline\noalign{\smallskip}
 $T\crit \, (\mathrm{K}) $     &  564.1   &   $P\crit\,(\mathrm{bar}) $ &  14.15 \\
 $v\crit \, (\mathrm{m^3/kg})$ &  0.00525  &   $ c_{v_\infty}/R$ &  57.69 \\
\noalign{\smallskip}\hline\noalign{\smallskip}
  \hspace{12mm}  & \textsf{Up-stream}\hspace{3mm} &  & \textsf{Down-stream} \hspace{2mm}  \\
   $P/P\crit$  &  0.704  & & 1.160  \\
   $v/v\crit$  &  3.000   & & 1.236  \\
   $T/T\crit$  &  1.037  & & 1.057  \\
   $M$         &  2.000  & & 2.873  \\
   $\Gamma$    &  0.667  & & 0.279  \\
\noalign{\smallskip}\hline
\end{tabular}
\end{table}

\begin{figure*}
\includegraphics[width=\hsize]{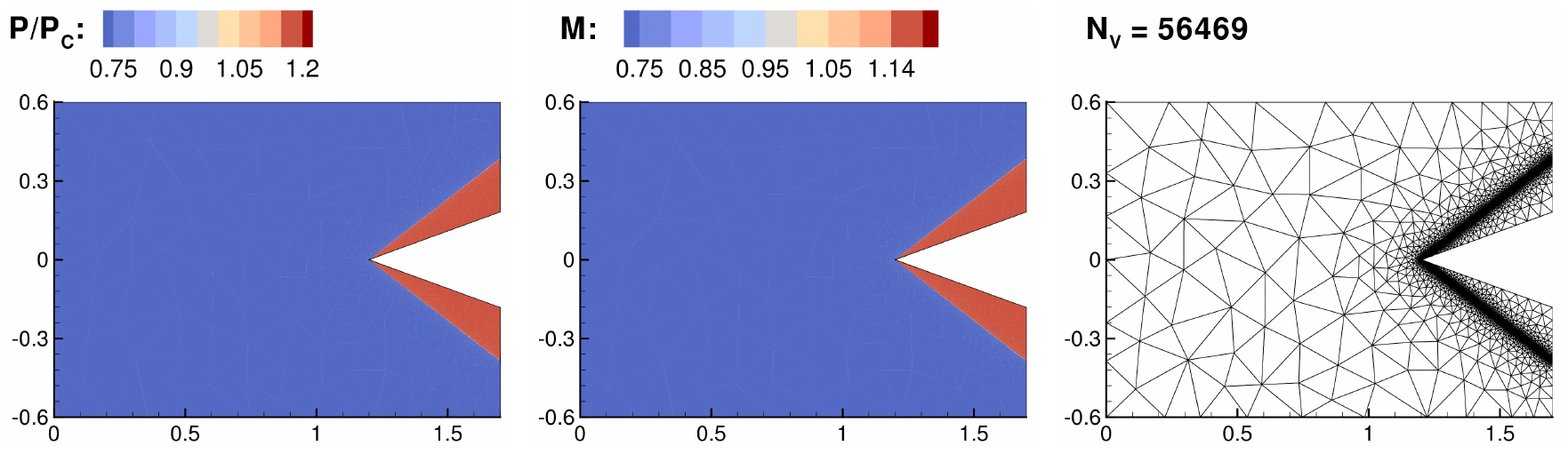}
\caption{Steady simulation of the non\hyp{}ideal oblique shock: pressure, Mach contour plots, and grid after five adaptation cycles. Pressure is scaled with respect to the critical value.}
\label{fOS:steady}
\end{figure*}

\begin{figure}
\includegraphics[width=\hsize]{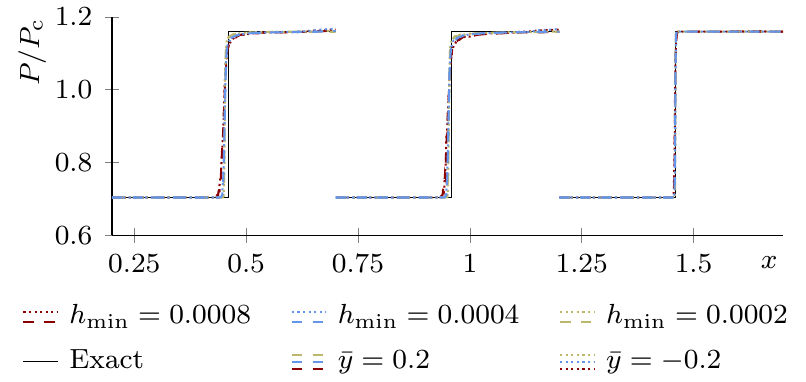}\\
\includegraphics[width=\hsize]{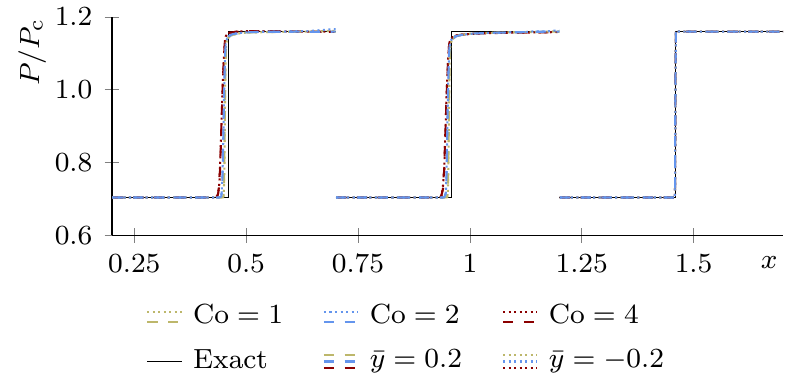}
\caption{Steady and unsteady simulations of the non\hyp{}ideal oblique shock. Top: different minimal edge sizes at $\mathrm{Co}=1$. Bottom: $h_\mathrm{min}=0.0002$ at different Courant numbers. The pressure, scaled with respect to the critical value, is extracted at two locations $\bar{y}=\pm 0.2$ and compared to the analytical one~\cite{Gori2017}, at three different times: when the wedge displacement is $\Delta x =0$ (the results of the steady simulation), $\Delta x =-0.5$, $\Delta x =-1$. Only the last portion of the pressure is plotted (for $x> x_\mathrm{R}-0.5$, with $x_\mathrm{R}$ the position of the right boundary).}
\label{fOS:profile}
\end{figure}

\begin{figure*}
\includegraphics[width=\hsize]{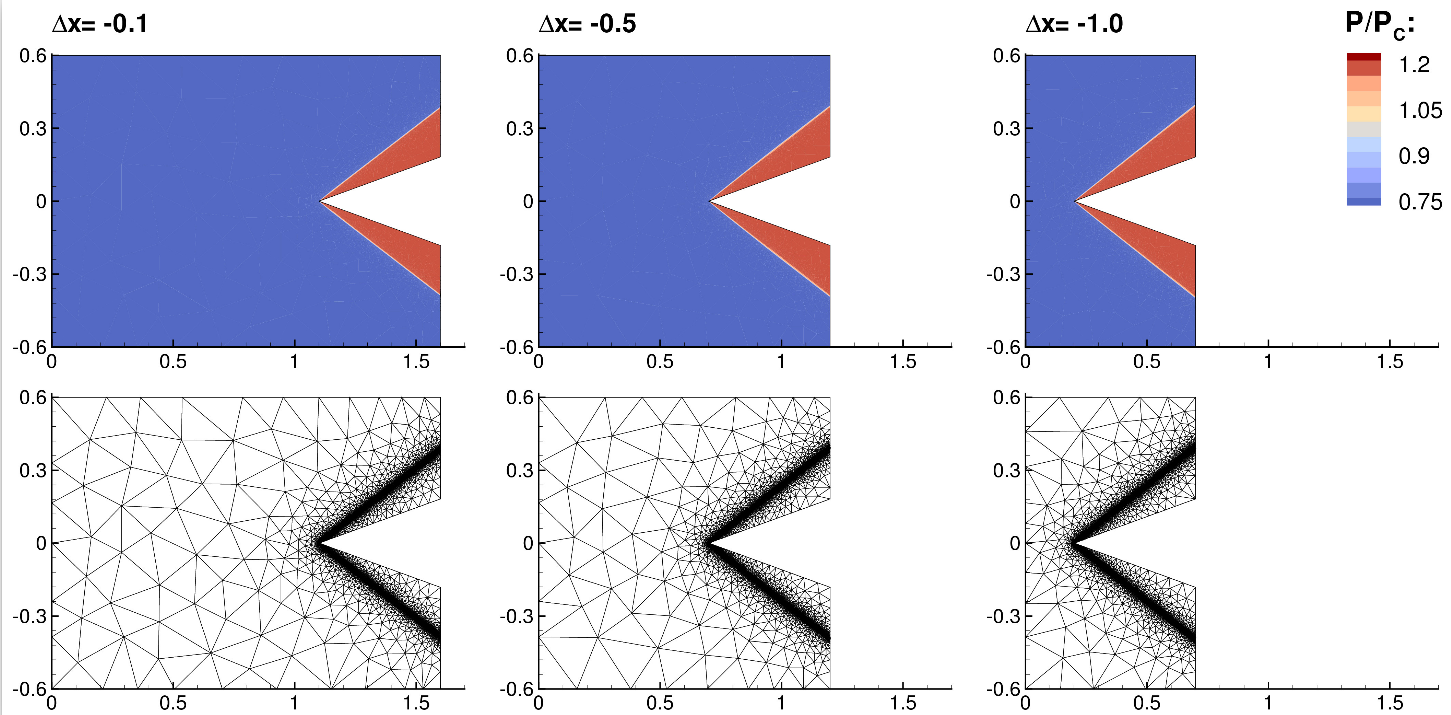}
\caption{Unsteady simulation of the non\hyp{}ideal oblique shock, with $h_\mathrm{min}=0.0002$ and $\mathrm{Co}=2$. Pressure contour plot and grid are shown at three different times, namely when the wedge displacement is $\Delta x =-0.1$, $\Delta x =-0.5$, $\Delta x =-1$.}
\label{fOS:contour}
\end{figure*}

We selected a specific configuration that highlights the potential non\hyp{}ideal effects in oblique shock waves.
As shown in~\cite{Gori2017}, when $0<\Gamma<1$, the non\hyp{}monotone variation of the speed of sound along isentropic expansions can result in an increase of the flow Mach number across oblique shock waves, contrary to what is predicted by the perfect gas model.
Here, we reproduce one of the tests described in~\cite{Gori2017} to compare our results with the analytical and numerical solution presented therein.
The working fluid is the siloxane $\mathrm{MDM}$\footnote{%
$\mathrm{MDM}$ is the acronym for the octamethyltrisiloxane,
whose chemical formula is \ensuremath{\mathrm{C_8 H_{24} O_2 Si_3}}},
and the test conditions are detailed in Table~\ref{t1:mdm}.
As in the reference, we adopt the polytropic van der Waals model. Further information about the implementation of this thermodynamic model in the flow solver can be found in~\cite{Re2017jet}.

The domain extends from $x=0$ to $x=1.7$ and from $y=-0.6$ to $y=0.6$ (assuming a unitary dimensional reference length of $1~\mathrm{m}$). The leading edge of the wedge is at $x=1.2$, $y=0$.
Although the domain is symmetrical with respect to the line $y=0$, it is discretized by an unstructured grid, so the symmetry is lost from the numerical point of view.
Inflow boundary conditions for density, velocity, and pressure are enforced on the right end\hyp{}wall, while the far\hyp{}field and the outflow are modeled by non\hyp{}reflecting boundary conditions that, where possible, impose only pressure and density.
Starting from an initial coarse grid, initialized everywhere with the up\hyp{}stream conditions, three adaptation strategies are performed by imposing three different minimal edge sizes: $h_\mathrm{min}=\left\lbrace 0.0008, 0.0004, 0.0002\right\rbrace$.
In all cases, an isotropic adaptation criterion is built by blending together the Hessian of the pressure and the gradient of the Mach number. Five adaptation cycles (\textit{Prediction}, \textit{Mesh adaptation}, \textit{Solution}) are performed, and all strategies successfully delivered  a refinement of the shock region, up to the minimum prescribed edge size.
The results obtained in the steady simulations with $h_\mathrm{min}=0.0002$ are shown in Fig.~\ref{fOS:steady}. 
Remarkably, the post\hyp{}shock Mach number is larger than the pre\hyp{}shock one, thus pointing out the non\hyp{}ideal nature of the shock.
A comparison against the analytical solution is given in Fig.~\ref{fOS:profile}:
an excellent agreement with the analytical solution is achieved.

The steady solutions are then used to initialize the unsteady simulations. The free stream velocity is subtracted from the solution and a displacement of $\Delta x =-1$ at $M=2$ is imposed to the wedge.
With respect to the three different minimal edge sizes, the whole simulation time is divided in $N_\mathrm{T}=\left\lbrace 1250, 2500, 5000 \right\rbrace$ steps in order to have a Courant number $\mathrm{Co}=1$.
The top picture in Fig.~\ref{fOS:profile} shows that the initial pressure profile (obtained in the steady simulation) is conserved well during the unsteady computations.
As expected, the size $h_\mathrm{min}=0.0002$ allows to achieve the best results, especially in terms of the post\hyp{}shock state. The pressure is slightly over\hyp{}predicted if one uses a lower resolution grid.
The symmetry of the problem is recovered in the numerical results, since no differences between the profiles extracted from the lower and the upper part can be appreciated.
Further simulations are performed with the highest grid resolution at $\mathrm{Co}=\left\lbrace 2, 4 \right\rbrace$, to assess the validity of the adaptive unsteady scheme also at higher Courant numbers.
The results, plotted in Figs.~\ref{fOS:profile}(bottom) and \ref{fOS:contour}, agree well with the analytical ones, and the shock resolution is deteriorated only to a small degree by the high Courant number.

\subsection{Unsteady simulations of a conical shock}
\begin{figure*}
\includegraphics[width=\hsize]{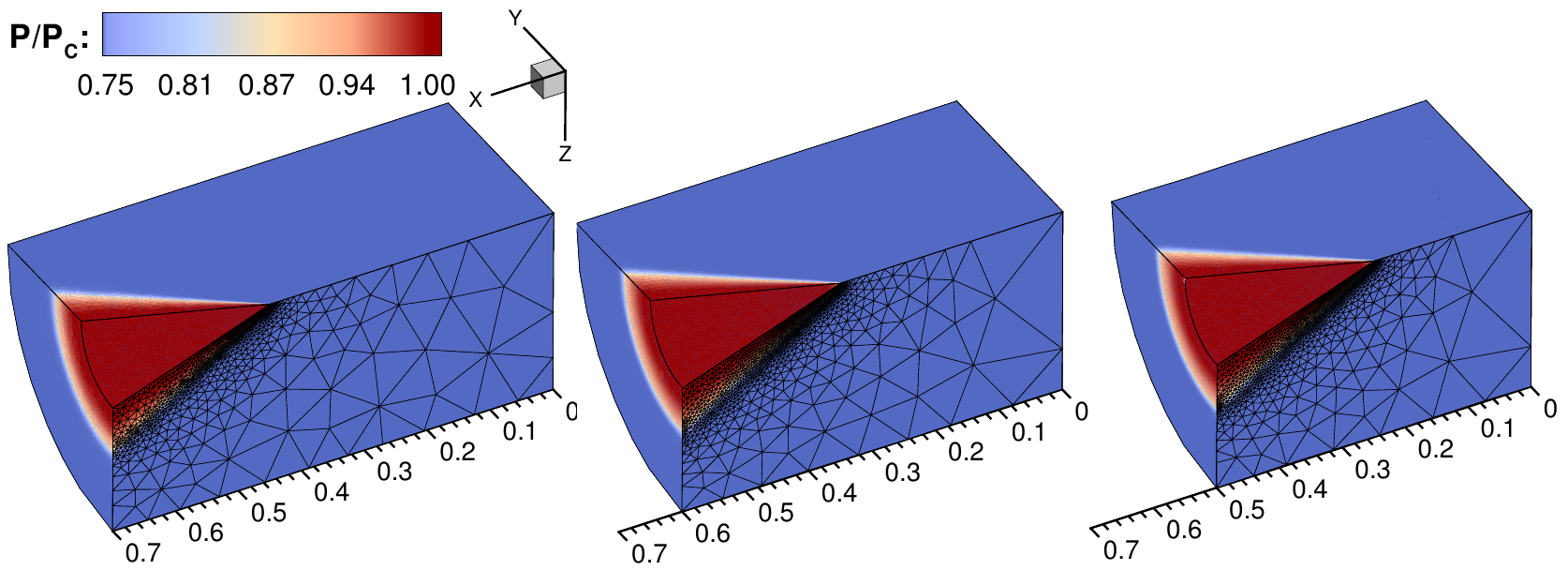}
\caption{Steady and unsteady simulations  of the non\hyp{}ideal conical shock, with $h_\mathrm{min}=0.0002$ and $\mathrm{Co}=2$. Pressure contour plot and the grid over the $y=0$ plane are shown at the beginning of the unsteady simulation, i.e., the results of the steady one (left),
when the cone displacement is $\Delta x =-0.1$ (in the middle), and when $\Delta x =-0.2$ (right).}
\label{fC3:3d}
\end{figure*}

\begin{figure}
\includegraphics[width=\hsize]{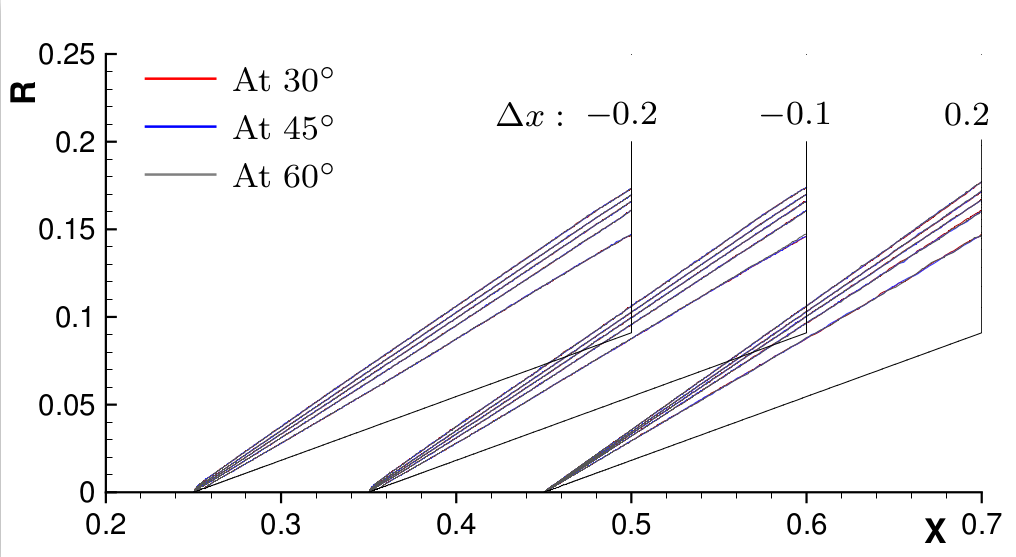}
\caption{Steady and unsteady simulations of the non\hyp{}ideal conical shock. Contour lines of the pressure over the planes sliced at $30^\circ$, $45^\circ$, and $60^\circ$ extracted at three different times, and plotted in the $x$-$R$ plane, where $R=\sqrt{y^2 + z^2}$ is the radial distance. The black lines represent the edges of the plane.
Five contour lines are plotted, which are  $P/P_\mathrm{c}=\left\lbrace 0.75, 0.8, 0.85, 0.9, 0.95 \right\rbrace$.
}
\label{fC3:contour}
\end{figure}

A test similar to the one described in the previous subsection is now presented to assess the validity of the proposed method in 3D. The geometry of the test can be viewed as the solid of revolution obtained by rotating half of the domain of the wedge test through 90 degrees about the $x$-axis. The starting wedge geometry used for the rotation is smaller than the one used in Subsec.~\ref{ssec:oblique}, and it extends from $x=0$ to $x=0.7$ and from $y=0$ to $y=0.3$.
In the resulting three\hyp{}dimensional domain, the far\hyp{}field is represented by a quarter of a cylindrical surface, and the solid body by a quarter of a conical one. The lateral planes at $y=0$ and at $z=0$ are modeled as solid walls to respect the symmetry. For the inflow and the outflow, the same non\hyp{}reflection conditions used in the oblique shock test are enforced.

An adaptive steady simulation with the free\hyp{}stream conditions given in Tab.~\ref{t3:state} is first performed.
Even though the upstream conditions are the same, the flow state downstream of the conical shock differs from the one downstream of the oblique shock, as is well known. No analytical solution of the conical shock waves with van der Waals EoS is available.

The steady results are then used to generate the initial solution for the unsteady computation,  which starts from a grid made of 150633 nodes and 882581 elements, with a minimum edge size of $h_\mathrm{min}=0.0002$. A displacement of $\Delta x = -0.2$ is imposed to the cone and a Courant number of $\mathrm{Co}=2$ is enforced.
The results are shown in Fig.~\ref{fC3:3d}. We can observe that the flow field is well conserved during the displacement thanks to the fine grid spacing obtained near the shock as a result of mesh adaptation.
Figure~\ref{fC3:contour} confirms the good agreement between the unsteady and the steady pressure fields. Moreover, three planes are sliced at different azimuth angles, and the resulting pressure contours are plotted in the $x$-$R$ plane, where $R=\sqrt{y^2 + z^2}$. An almost perfect overlap between contour plots over the three planes is achieved, which confirms the capability of the proposed approach to correctly recover the cylindrical symmetry of the flow field.
Mesh adaptation plays a crucial role to achieve this result, because only a proper refinement of the grid regions where the flow variations are significant can overcome the asymmetry of the unstructured grid.

\section{Numerical investigation of unsteady piston problems}
\label{sec:results}

\begin{figure}
\centering
\includegraphics[width=0.9\hsize]{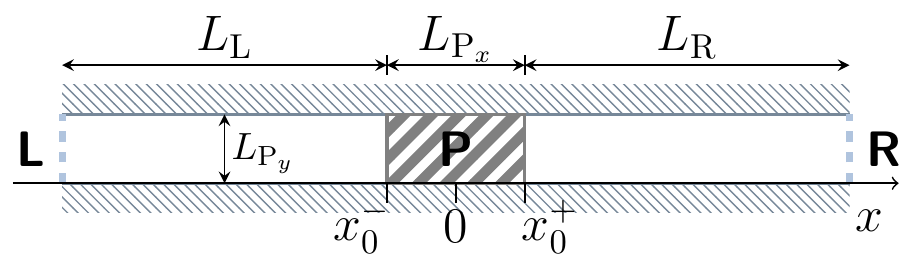}
\caption{Sketch of the reference piston problem. The piston (\textsf{\textbf{P}}) moves inside a tube, and no gap is modeled between the piston and the tube walls. The letters \textsf{\textbf{L}} and \textsf{\textbf{R}} indicate respectively the right and the left side of the tube, which, according to the test case, can be open-ends or solid walls.
The $x$-axis is placed along the lower side of the tube and $x=0$ is in the middle of the piston.
The length and the height of the piston are $L_{\mathrm{P}_x} = 0.1~\mathrm{m}$ and $L_{\mathrm{P}_y} = 0.05~\mathrm{m}$, respectively. The initial position of the left and right piston faces are $x_0^-= -0.05~\mathrm{m}$ and $x_0^+= 0.05~\mathrm{m}$.}
\label{f:piston}       
\end{figure}

The adaptive computational procedure described in the previous sections is here further assessed through the simulations of different piston problems, in 2D and 3D.
The piston problem is a standard gasdynamics test which presents diverse numerical challenges.
It models the motion of a piston inside a tube with no gap between the tube and the piston itself.
If the piston moves towards the right, the fluid lying in the right part of the tube is compressed, the compression wave moves away from the piston and travels rightwards. Correspondingly, in the left part of the tube an expansion wave is generated and travels toward the left.
While these nonlinear waves propagate through the tube, their waveform is distorted.
If $\Gamma>0$, the compressive waves steepen, possibly forming a classical compression shock, and the rarefaction waves spread out.
On the contrary, if $\Gamma<0$, the compressive part of the wave spreads over a longer space as it moves, while the expansion part steepens and it may possibly form a rarefaction shock, the most representative flow structure of the non\hyp{}classical regime.
If the tube is closed, the waves are reflected at the tube ends and interact with the incoming waves, originating a complex flow pattern which contains compression and expansion waves of different intensities, shocks, and contact discontinuities.
In addition, the intensity and the type of the perturbation waves can be roughly controlled by the piston motion, thus different tests can be performed.
A flow field encompassing simultaneously all these flow features represent a valuable, challenging test to assess the proposed method, since the finite\hyp{}volume solver is edge\hyp{}based and the numerical simulations are performed over unstructured grids, so they are inherently multidimensional. Rather, this feature provides an additional opportunity to assess the validity of the proposed method.

The geometry of the reference piston problem is shown in Fig.~\ref{f:piston}.
The tube has different lengths depending on performed tests.
For 3D simulations, a square section is assumed, i.e., $L_{\mathrm{P}_z} = L_{\mathrm{P}_y}$.
The tube is filled with the linear siloxane \mdqm\footnote{%
\mdqm\ is the acronym for the tetradecamethylhexasiloxane,
whose chemical formula is \ensuremath{\mathrm{C_14 H_{42} O_5 Si_6}}},
which is a high-molecular complexity fluid that presents a quite extended $\Gamma<1$ region and a small one at $\Gamma<0$.
Table~\ref{t2:md4m} summarized the relevant thermo\hyp{}physical properties of \mdqm.
As expected, the compressibility factor at the critical point is different from the one predicted by the Peng--Robinson EoS. Indeed, the specific volume computed through the adopted cubic thermodynamic model slightly deviates from the value reported in Tab.~\ref{t2:md4m}. 

\begin{figure*}\sidecaption
\includegraphics[scale=1]{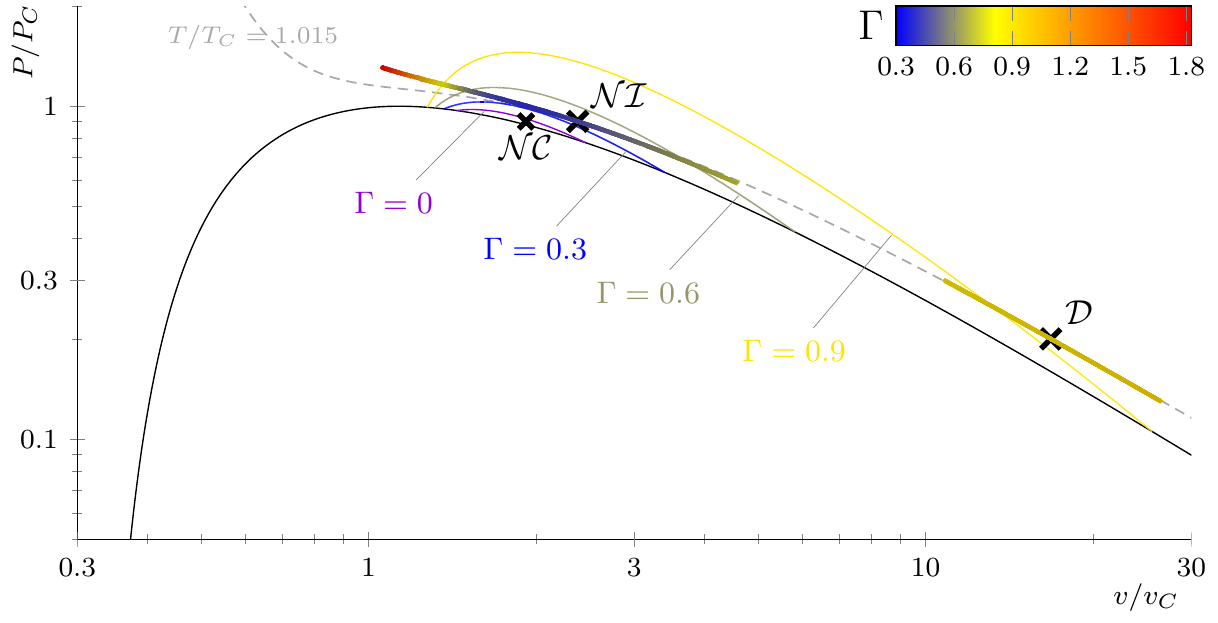}
\caption{Description of the tests.
The thick crosses~\markCaption{mark=x, ultra thick, mark size=4pt} display the initial states of the fluid at rest in the tube. For \testA\ and \testB, selected along the same isotherm at $T_0=1.015 T\crit$ (dashed line ~\line{DarkGray, no marks, densely dashed, line width=0.5pt}),
the flow field resulting after one period of oscillation in the infinite-length tube is plotted and colored according to the value of $\Gamma$. 
In addition, the VLE curve (black line~\line{black}) and four curves at constant $\Gamma$ are displayed.
The coordinates are scaled with respect to the critical values reported in Tab.~\ref{t2:md4m}. Note that the critical point coordinates computed by the Peng-robinson model are not unity, but $v_c^{\mathsf{PR}}/v_c=1.149$.
}
\label{f:vleAB}
\end{figure*}

Initially, the fluid inside the tube is at rest (labeled with subscript 0).
Three different initial states are used to investigate peculiar gasdynamic behavior of the fluid: 
the initial conditions for the dilute (\testA) and the NICFD (\testB) regimes are selected along an isotherm above the critical one, while the initial conditions for the non\hyp{}classical case (\testC) are selected along an isotherm below the critical one.
The initial states in all tests are detailed in Tab.~\ref{t3:state} and they are displayed on the $P-v$ plane, along with the VLE curve, in Fig.~\ref{f:vleAB}.

In this work, two different types of motion are simulated: a harmonic motion which transfers to the fluid in contact with the piston a sinusoidal perturbation wave, and an impulsive start, which generates instantaneously a shock wave and a rarefaction fan which travel outwards.
In the harmonic motion, the position of the piston is set as
$ x(t) = x_0(t) + A \cos (2 \pi \,f \, t) $,
where $A$ is the amplitude and $f$ the frequency ($f=30~\mathrm{Hz}$ for all tests).
In the other case, the piston impulsively acquires a constant positive velocity $V_\mathrm{P}= \vert A \pi \,f\vert$, that is half of the maximum velocity experienced by the piston during the harmonic motion.
The piston motion represents an additional challenge for the CFD software, as it requires dealing with large deformations, especially in the simulation of the impulsive start, where the piston displacement amounts to more than the $40 \%$ of the length of the domain.

The harmonic motion is first simulated in a tube with infinite\hyp{}length to assess the capability of reproducing both the expansion and compressive waves generated by the motion and to observe their distortion. Then, the simulation is repeated considering a closed tube, to assess also the capability to correctly model the interactions with solid boundaries and other waves.

In the following subsections, the exposition and discussion of the results starts from the 2D tests in the NICFD region, with the harmonic motions of the piston in a infinite\hyp{}length tube  and in a finite\hyp{}length tube, followed by the impulsive start in a closed tube. Then, results concerning the non-classical region are shown for the harmonic motion of the piston in an infinite\hyp{}length tube, with two different amplitudes.
Finally, some 3D results are displayed to show the capability of tackling 3D problems.
In all figures of this section, a reference length of $1~\mathrm{m}$ is used to make all distances dimensionless.

\begin{table}
\caption{Thermodynamic properties for \mdqm. The molecular mass $M_\mathrm{m}$, the boiling temperature $\Tb$, the critical temperature and pressure are those included in the NIST software {REFPROP}~\cite{Refprop}. The critical volume is given by the 12-parameter EOS~\cite{Colonna2006a} implemented in the same software.
The isobaric specific heat capacity in the ideal gas limit $c_{p_\infty}$ is computed through the polynomial expression given in~\cite{Colonna2006a}.
The acentric factor is approximated by Eq.~\eqref{e:acentricFactor}}
\label{t2:md4m} 
\begin{tabular}{lllll}
\hline\noalign{\smallskip}
 $M_\mathrm{m}\,(\mathrm{kg/mol})$   &  0.45899  &  &  $\Tb \, (\mathrm{K}) $  & 533.9\\
 $T\crit \, (\mathrm{K}) $     &  653.20   &  &  $P\crit\,(\mathrm{bar}) $ &  8.7747 \\[1ex]
 $v\crit \, (\mathrm{m^3/kmol})$ &  1.7309  &  &  $Z\crit$   & 0.2797 \\
$ c_{p_\infty}/R$ &  115.99  &   &  $\omega$  & $0.7981$\\
\noalign{\smallskip}\hline
\end{tabular}
\end{table}

\begin{table}
\caption{Initial thermodynamic state for the three tests.}
\label{t3:state} 
\begin{tabular}{llll}
\hline\noalign{\smallskip}
  \hspace{12mm}  & \testA \hspace{3mm} &  \testB \hspace{2mm} & \testC \\
\noalign{\smallskip}\hline\noalign{\smallskip}
 \textsf{Regime}  & dilute &  non-ideal  & non-classical \\
 $T_0/T\crit$ &  1.015   &   1.015  &  0.9955  \\
 $P_0/P\crit$ &  0.2     &    0.9   &  0.9 \\
 $v_0/v\crit$ &  16.780  &   2.374  &  1.918 \\
 $Z_0      $ &  0.9257  &   0.5886 &  0.4861  \\
 $\Gamma_0 $ &  0.9306  &   0.4516 &  -0.0064  \\
 $c_0\,(\mathrm{m/s})$ & 101.9 & 61.9 & 45.2 \\
\noalign{\smallskip}\hline
\end{tabular}
\end{table}

\subsection{2D oscillating piston in an infinite\hyp{}length tube: dilute and non\hyp{}ideal conditions}
\label{ssec:infinite2D}
\begin{figure*}\centering
\includegraphics[width=\hsize]{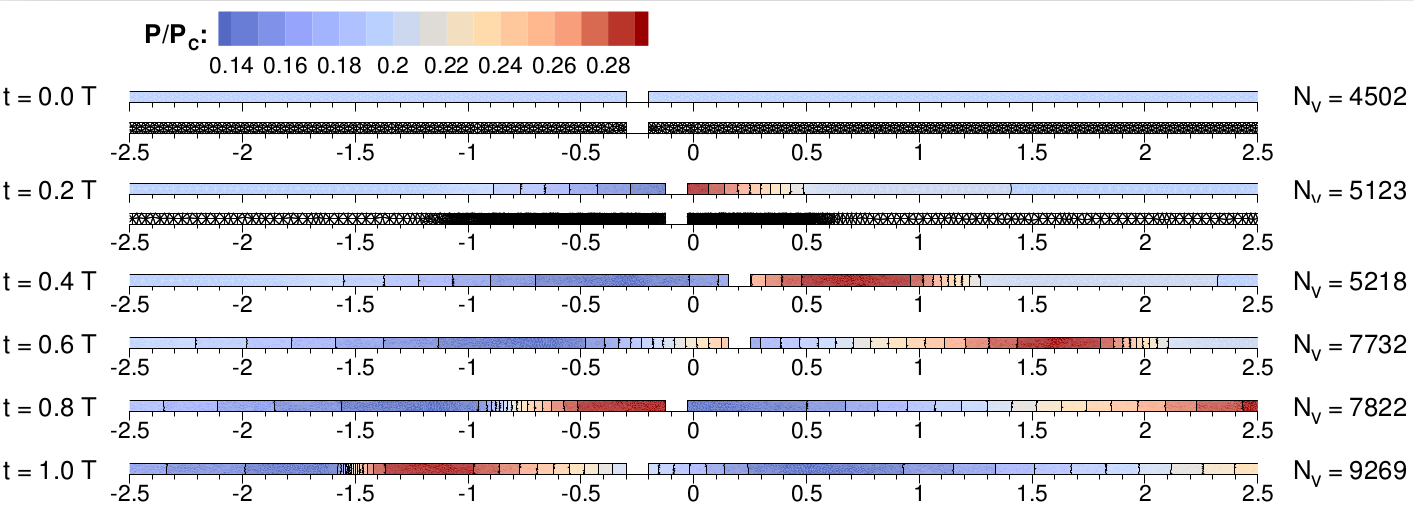}\\[2ex]
\includegraphics[scale=1.1]{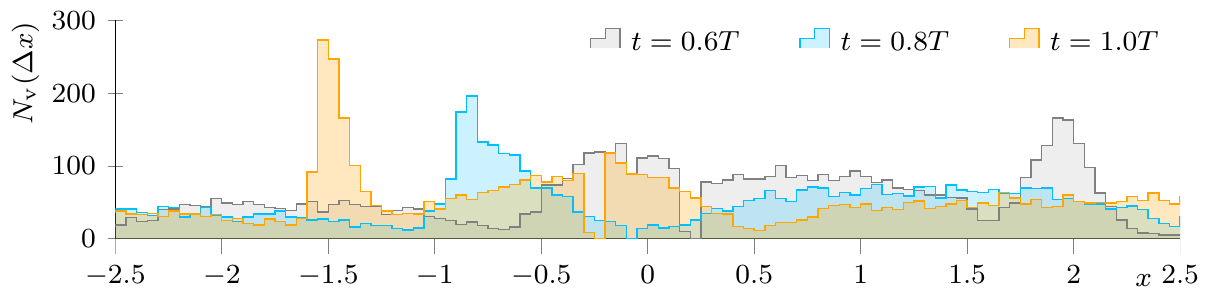}
\caption{Oscillating piston in an infinite\hyp{}length tube, \testA:  central portion of the domain, near the piston.
Top: pressure contour plots during the first period, initial grid, and grid at $t=0.2 T$. The motion of the piston is clearly visible (initial position $x_0^- = -0.3$ and $x_0^+= -0.2$).
Bottom: histogram of the number of grid nodes along $x$-axis, bin width $\Delta x=0.05$.
A comparison between the grid histogram and the pressure plot highlights the effect of mesh adaptation. For instance, by looking at the right part, at $t=0.8 T$ the almost equi-spatially distributed contour lines are reflected by the almost uniform grid density.
In the left part of the domain, the mesh adaptation leads to an increase of the grid point density as the compressive waves coalesce.
}
\label{fOB:contourID}
\end{figure*}

\begin{figure*}
\centering
\includegraphics[width=\hsize]{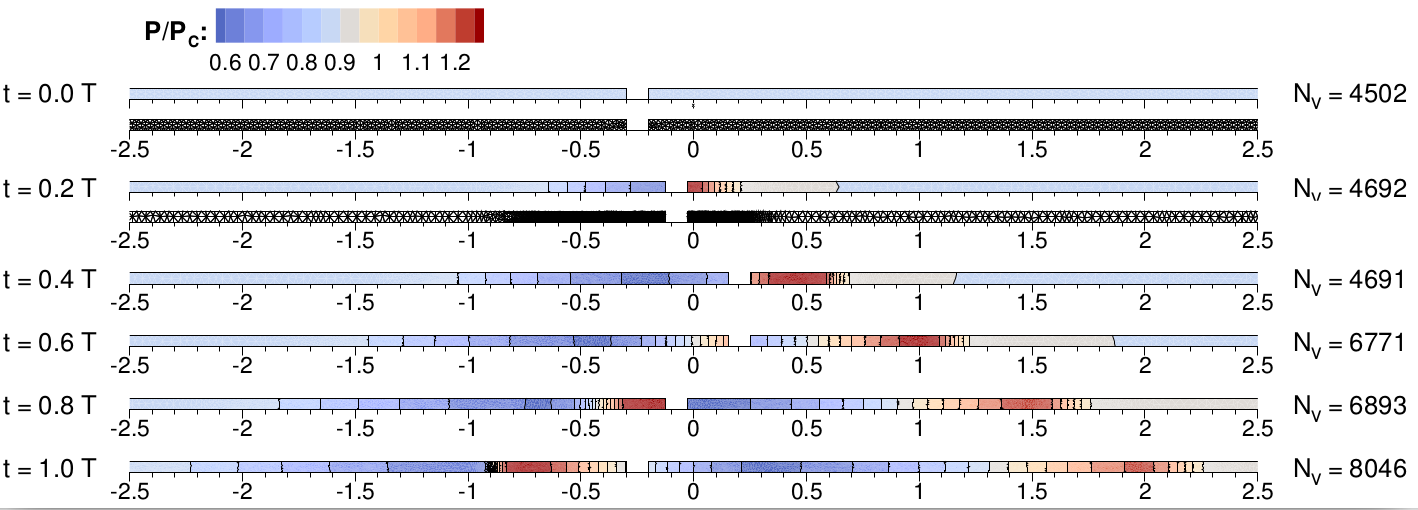}\\[2ex]
\includegraphics[scale=1.1]{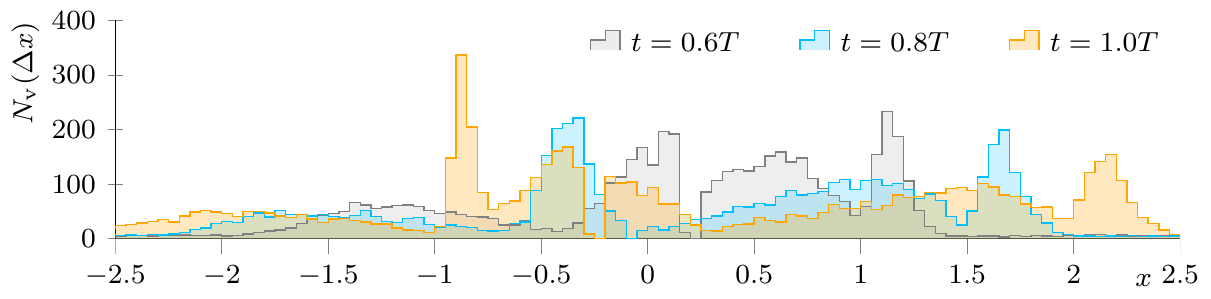}
\caption{Oscillating piston in an infinite\hyp{}length tube, \testB:  central portion of the domain, near the piston.
Top: pressure contour plots during the first period, initial grid, and grid at $t=0.2T$. The motion of the piston is clearly visible (initial position $x_0^- = -0.3$ and $x_0^+= -0.2$).
Bottom: histogram of the number of grid nodes along $x$-axis, bin width $\Delta x=0.05$.
A comparison between the grid histogram and the pressure plot highlights the effect of mesh adaptation. The peaks in the right part of the domain correspond to the compression that reduces its strength as it travels outwards. The height of the peaks in the left part increases from $t=0.6T$ to $t=T$, leading to the strongest  refinement in correspondence of the shock wave (around $x=-0.9$ at $t=T$).
}
\label{fOB:contourNICFD}
\end{figure*}

The first test case consists in the harmonic motion of the piston in an infinite\hyp{}length 2D tube, with an amplitude $A=-0.25~\mathrm{m}$.
In these simulations, the focus is on the compressive and expansive waves generated by the harmonic motion, so we have limited the influence of the boundary on the flow field by modeling the left and right ends as normal inlets/outlets, and by imposing the tube lengths $L_\mathrm{L}=L_\mathrm{R}=10~\mathrm{m}$.
The initial grid is composed of $6998$ triangles and $4502$ nodes and three oscillation periods are simulated.

The piston starts moving towards the right.
Its velocity progressively increases during the first quarter of the oscillation period, $0<t<0.25T$, until $V_{\mathrm{P},\mathsf{max}} = -2 \pi f \, A$. During this interval, a compressive wave is generated on the right side of the piston and moves towards the right end of the tube.
In the next quarter of the period $0.25T<t<0.5T$, the piston still moves towards the right but its speed reduces ($V_\mathrm{P}=0$ at $t=0.5T$). This gives rise to an expansion on the right section and a compression on the left part, which continue also in the next quarter of period.
At $t=0.75T$ the velocity of the piston is minimum, $V_{\mathrm{P},\mathsf{min}} =2 \pi f \, A$ (same magnitude of $V_{\mathrm{P},\mathsf{max}}$ but negative),  then, it increases leading to an expansion on the left side and a compression on the right side, similarly to the first quarter.
The flow fields at different times during the first oscillation period are shown in Fig.~\ref{fOB:contourID} for \testA\ and in Fig.~\ref{fOB:contourNICFD} for \testB.
The solutions at $t=T$ are shown also in the $P$-$v$ plane in Fig.~\ref{f:vleAB}, from which the significant variation of $\Gamma$ in \testB\ can be appreciated.

During the next periods, the flow field evolves in a similar way.
Nevertheless, unlike the ones occurring in the right part at $0<t<0.25T$, the following compressions are generated by the largest variation of the piston velocity, i.e., from the minimum to the maximum value, or vice versa. Therefore, these waves may coalesce together generating a shock far from the piston side.

\begin{figure}
  \includegraphics[width=\hsize]{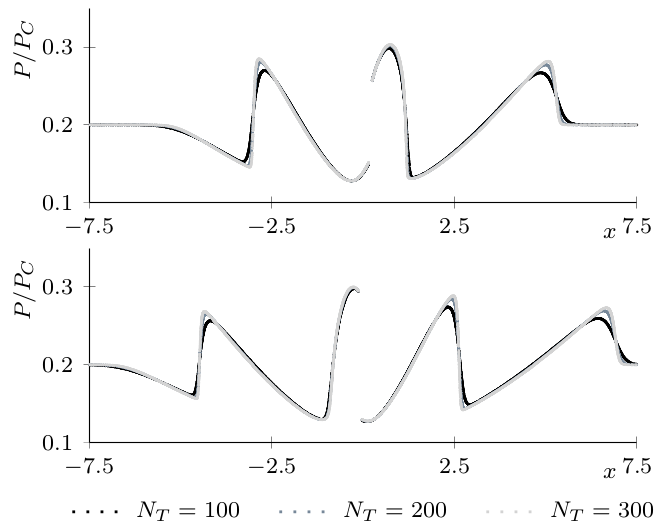}
\caption{Oscillating piston in an infinite\hyp{}length tube, \testA: pressure at time $t=1.4~T$ (top) and at $t=1.8~T$(bottom), obtained with three different numbers of time steps per period: $N_T=100$ in black~\markCaption{mark=*, mark size=1pt,black}~,
$N_T=200$ in dark gray~\markCaption{mark=*, mark size=1pt,LightSlateGrey}
and $N_T=300$ in light gray~\markCaption{mark=*, mark size=1pt,LightGrey}~.
Each point corresponds to a grid node.}
\label{fOB:cflID}       
\end{figure}

\begin{figure}
  \includegraphics[width=\hsize]{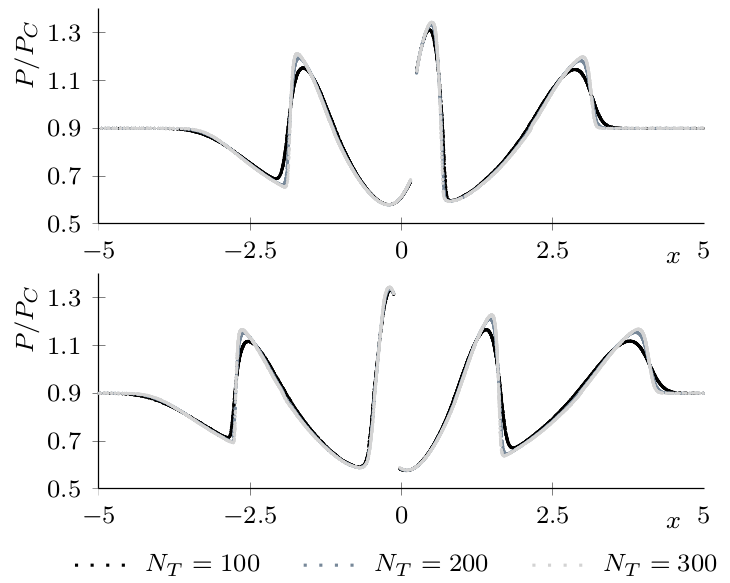}
\caption{Oscillating piston in an infinite\hyp{}length tube, \testB: pressure at time $t=1.4~T$ (top) and at $t=1.8~T$(bottom), obtained with three different numbers of time steps per period: $N_T=100$ in black~\markCaption{mark=*, mark size=1pt,black}~,
$N_T=200$ in dark gray~\markCaption{mark=*, mark size=1pt,LightSlateGrey}
and $N_T=300$ in light gray~\markCaption{mark=*, mark size=1pt,LightGrey}~.
Each point corresponds to a grid node.}
\label{fOB:cflNICFD}       
\end{figure}

Simulations with $N_T=\left\lbrace 100, 200, 300 \right\rbrace$ steps per period are carried out to estimate the minimum number that does not impair the solution accuracy.
Figures~\ref{fOB:cflID} and \ref{fOB:cflNICFD} show the pressure profile obtained with at $t=1.4 T$ and $t=1.8 T$ for \testA\ and \testB, respectively. For both cases, we can observe that the largest time step leads to a loss of accuracy in proximity of the pressure maxima and minima, while no significant deviations can be noted for the other two time steps. Therefore, $N_T=200$ steps per period are used in all following simulations.
Although the resulting plots appear as single lines, Figs.~\ref{fOB:cflID} and \ref{fOB:cflNICFD} are in fact scatters of all points of the unstructured grids.
The one\hyp{}dimensional behavior of the flow is therefore perfectly recovered in both tests, and this gives the first, partial, assessment of the validity of the proposed scheme.
Small symmetry disturbances that can be detected in the contour plots of Figs.~\ref{fOB:contourID} and \ref{fOB:contourNICFD} are not present in the pressure profiles, and they do not affect the symmetry of the results. In our opinion, they could be ineffective consequences of the unstructured grids, which do not undermine the validity of the proposed approach.

\begin{figure*}
  \includegraphics[width=\hsize]{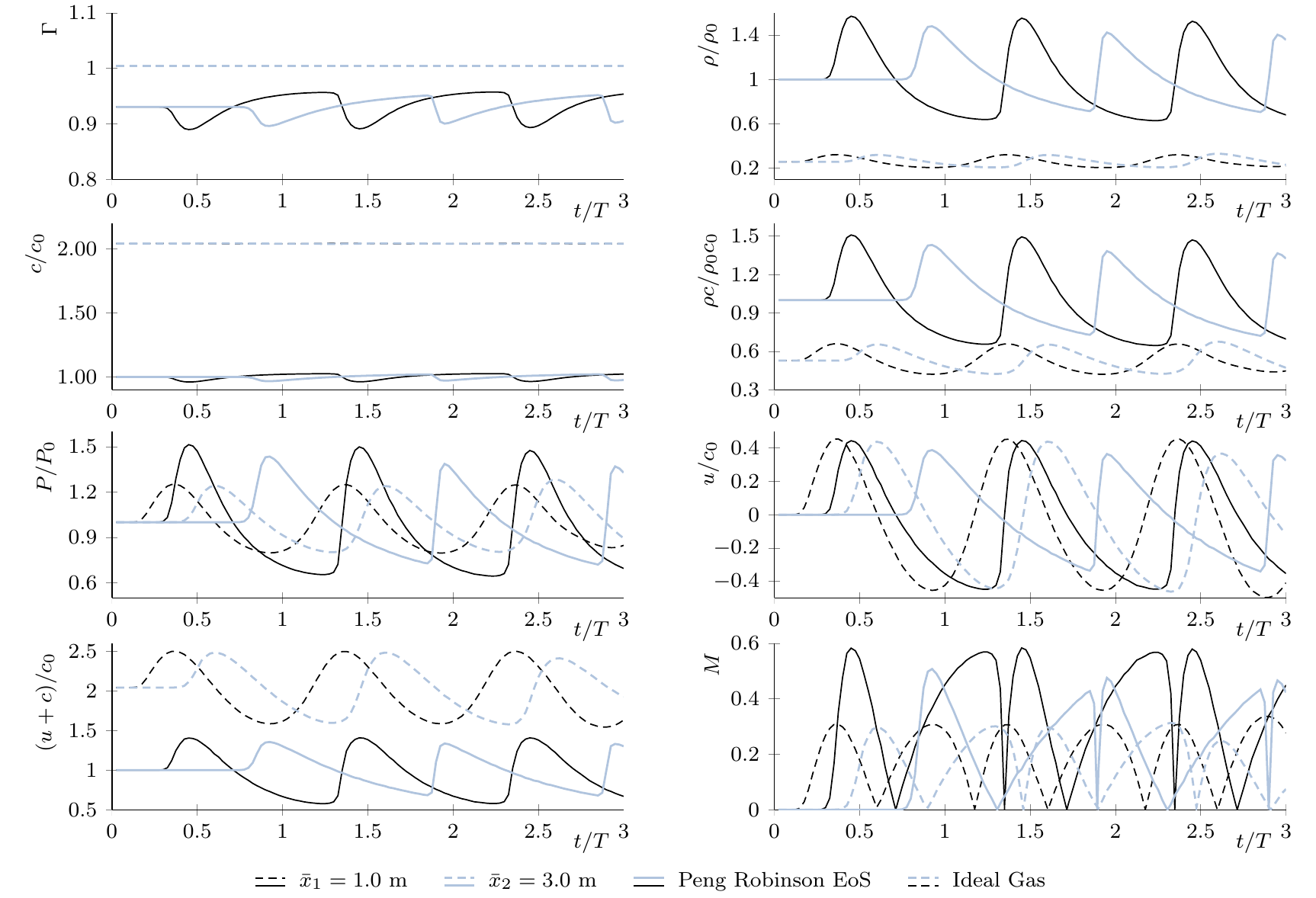}
\caption{Oscillating piston in an infinite\hyp{}length tube, \testA: temporal evolution of eight thermodynamic quantities at two different locations in the right part of the tube, one near the piston ($\bar{x}_1$, dark lines~\line{black}) and one farther ($\bar{x}_2$, light\hyp{}coloured lines~\line{LightSteelBlue}). From the first row to the last one: fundamental derivative, density, speed of sound, acoustic impedance, pressure, $x$-component of the fluid velocity, wave propagation speed, and Mach number. The results obtained with the polytropic Peng-Robinson EoS (solid lines) and the PIG model (dashed lines) are compared.
Pressure, density, and velocities are scaled with respect to the initial values $P_0$, $\rho_0$, and $c_0$ computed with the Polytropic Peng--Robinson model and reported in Tab.~\ref{t3:state}.
Note that, having kept fixed $P_0$ and $T_0$, the initial density and speed of sound computed through the ideal gas are different from those used to scale the values.
}
\label{fOB:probeID}       
\end{figure*}


In these tests, an anisotropic metric is built on the basis of the Hessian of the Mach number and two adaptation cycles are performed every time step. More precisely, the estimator is obtained dividing the Hessian by the gradient to reduce the predominance of the strongest phenomena~\cite{Re2017jet}.
The beneficial effects of mesh adaptation are clearly displayed by Figs.~\ref{fOB:contourID} and \ref{fOB:contourNICFD},
which, in addition to the pressure contours, show the histograms of the number of grid points along the $x$-axis, and two grids.
As perturbation waves propagate outwards, mesh adaptation affects a larger portion of the domain, as it can be seen from the grids at $t=0T$ and $t=0.2T$ (grids at the next times are not shown since the node density in some regions becomes so high that mesh elements are indistinguishable at this resolution).
Exploiting the one\hyp{}dimensional character of the flow, the grid density at three subsequent times is represented by the histogram of the number of grid nodes in fixed\hyp{}width bins. 
This type of graph allows to highlight the effect of mesh adaptation on a broad scale: 
the grid points are gathered where the flow field exhibits large variations, while the grid density is reduced in correspondence of almost uniform regions.
Grid coarsening is of paramount importance in unsteady simulations, since it allows limiting the number of grid nodes.
Mesh adaptation allows following flow variations as, for instance, occurs with the compression in the right part of the tube in Fig.~\ref{fOB:contourNICFD}: the peak in the histogram around $x=1.15$ at $t=0.6 T$ is progressively spread and moved towards the left, following the intensity and position of the compression.
Moreover, the grid density reflects the strength of the flow variations, as it can be clearly noticed by looking at the compressive waves and the corresponding histogram peaks in the left part of the tube in both tests. 
Therefore, the proposed approach is capable of linking the grid spacing with the unsteady flow features. 

A minor remark about Fig.~\ref{fOB:contourID} and \ref{fOB:contourNICFD} concerns the coarsening of the unperturbed regions of the initial grid, resulting, for example, in the different grid spacing for $x<-1.2$ between the grids at $t=0$ and $t=0.2T$ in Fig.~\ref{fOB:contourID}.
According to our experience, an extremely coarse initial grid may jeopardize the development of important flow features, so it is preferable to start with a barely coarse mesh and exploit mesh coarsening already in the first adaptation step, which is carried out after computing the solution in the \textit{Prediction} step, as explained in Subsec.~\ref{ssec:summary}. 
If, as in these tests, the maximum edge size allowed during the adaptation is different from the one of the initial grid, and the flow field at the beginning of the simulation embeds large uniform regions, the first adaptation step profoundly modifies the initial grid.
The regions that remain uniform during the next time steps are not affected by the following mesh adaptations.

\paragraph{Analysis of the flow behavior in dilute regime.}
To better analyze the flow behavior, the values of different flow quantities are extracted at fixed locations during the whole simulation time. For \testA, these quantities are shown in Fig.~\ref{fOB:probeID}.
The results obtained with the Polytropic Peng--Robinson (PPR) EoS are compared to the ones obtained with the Polytropic Ideal Gas (PIG) model, computed starting from the same $P_0$ and $T_0$.
Although the initial condition of this test lies in the dilute regime where the flow behavior is expected to differ only quantitatively from the ideal gas model, a first, general observation of Fig.~\ref{fOB:probeID} reveals that the differences between the two thermodynamic models are significant.
The most notable deviations affect the density and speed of sound profiles.
We can interpret these deviations as a measure of the non\hyp{}ideal gas effects.
During the simulation, the value of $\Gamma$ oscillates between $0.83$ and $0.96$, but the PIG model is not able to detect it, as it predicts a constant $\Gamma=1.004$.
This difference is extremely important, because it is reflected in the speed of sound.
Indeed, reminding Eq.~\eqref{e:Gamma},
a value of $\Gamma$ constant and close to unity results in a constant value of the speed of sound along an isentropic expansion, as shown in the profile for the PIG model. Conversely, for the PPR model, the shape of the profile of $c$ mirrors the one of $\Gamma$.

The differences on $c$ and $\rho$ affect the propagation of the discontinuities originated on the piston surface. For weak discontinuities, the acoustic theory links the pressure and velocity perturbation ($\delta P$ and $\delta u$, respectively) through the initial acoustic impedance $\rho_0 c_0$, namely
$ \delta P = \rho_0 c_0 \delta u$.
This relation suggests that, for the piston problem under investigation here where the only perturbation is the piston movement, if density and sound speed are inaccurate,
the same velocity perturbation generates a very different pressure perturbation.
For instance, defining the deviation on the pressure as $E_P = 100 (P^\mathsf{PR} - P^\mathsf{IG})/(P^\mathsf{PR}-P_0)$, the deviations on the maximum and minimum values are
\begin{description}
\item[at $\bar{x}_1=1.0$:] $\quad E_{P,\mathsf{max}} = 51\% , \qquad E_{P,\mathsf{min}} = 42\%$,
\item[at $\bar{x}_2=3.0$:] $\quad E_{P,\mathsf{max}} = 45\% , \qquad E_{P,\mathsf{min}} = 29\%$,
\end{description}
which show the failure of the PIG model to describe the flow behavior in this region, despite a value of $\Gamma$ near $1$.

A further difference that can be observed in the profiles concerns the shape.
In the PIG model, the perturbation is propagated from the piston surface through the tube with an almost constant shape, while according to the PPR model the initial sinusoidal shape loses its symmetry and the compression part becomes steeper.
This can be explained by watching the quantity $u+c$, which represents the wave speed in the right part of the piston. 
For a one-dimensional homentropic flow, the method of characteristics gives
\begin{equation}
\label{e:duc}
\mathrm{d} \, (u+c)=\frac{\Gamma}{\rho c} \mathrm{d} P \, .
\end{equation}
Thus, for the PIG model the higher value of $\Gamma$ and the lower value of the acoustic impedance generate much faster waves, which take shorter time to reach the same position.
Finally, for both models, being $\Gamma>0$ the compressive part of the wave becomes steeper, until a shock may form, generating a total pressure loss. This phenomenon is captured only by the PPR model, which predicts a faster increase of $u+c$. For this reason, the profiles at the location farther from the piston surface ($\bar{x}_2=3.0$) display different peak values, although these differences are limited since the shocks are quite weak.

For the sake of completeness, Fig.~\ref{fOB:probeID} reports also the Mach number profile, which embeds all the deviation explained above and clearly displays the loss of the initial sinusoidal shape as the maxima and the minima are not equally spaced.

\begin{figure*}
  \includegraphics[width=\hsize]{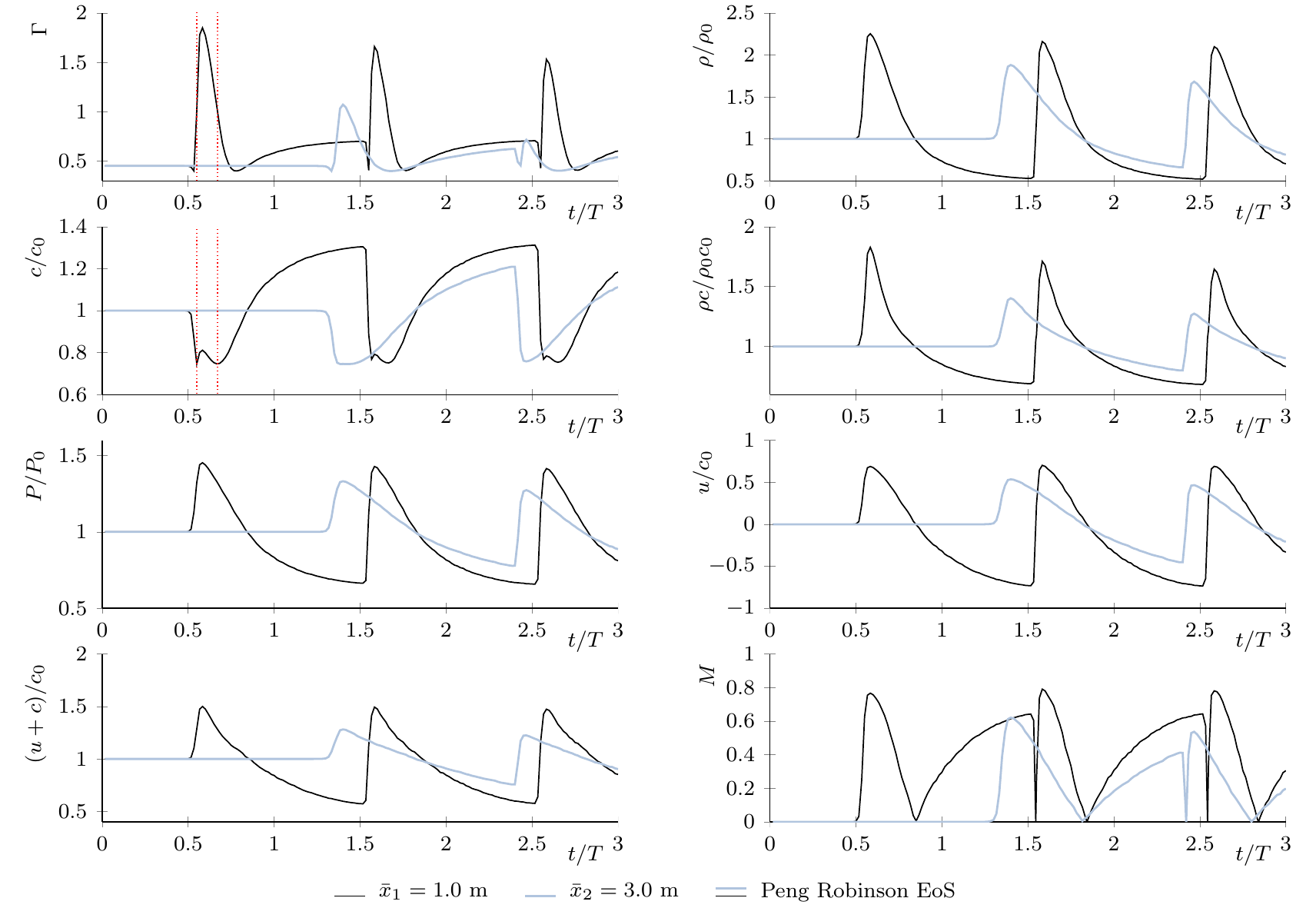}
\caption{Oscillating piston in an infinite\hyp{}length tube, \testB: temporal evolution of eight thermodynamic quantities at two different locations in the right part of the tube, one near the piston ($\bar{x}_1$, dark lines~\line{black}) and one farther ($\bar{x}_2$, light\hyp{}coloured lines~\line{LightSteelBlue}). From the first row to the last one: fundamental derivative, density, speed of sound, acoustic impedance, pressure, $x$-component of the fluid velocity, wave propagation speed, and Mach number. 
Pressure, density, and velocities are scaled with respect to the initial values $P_0$, $\rho_0$, and $c_0$ reported in Tab.~\ref{t3:state}.
The vertical dotted lines (\line{densely dotted, red}) enclose the first region where $\Gamma>1$, which occurs in the last part of the first compression and in the first part of the next expansion.
}
\label{fOB:probeNI}       
\end{figure*}

\paragraph{Analysis of the flow behavior in NICFD regime.}
The same quantities analyzed for \testA\ are shown in Fig.~\ref{fOB:probeNI} for \testB, which takes place in the core of the NICFD regime, as clearly displayed by Fig.~\ref{f:vleAB}.
The variations of the fundamental derivative $\Gamma$ profile are much more significant than the ones in \testA\ and the super-critical regime is also reached, in which the behavior of the flow is more similar to the liquid than the vapor one.
A peculiar phenomenon of the NICFD regime can be observed looking at the $c$ profile.
According to Eq.~\eqref{e:Gamma}, the variation of the speed of sound during an isentropic expansion is proportional to $\Gamma-1$. So, in the first part of the expansion where $\Gamma>1$, the speed of sound decreases, as occurs in ideal gas, while it increases in the second part, where $\Gamma<1$.
Nevertheless, the increase is not sufficiently strong to result in a non-monotone Mach profile.
Watching at the other quantities, a behavior similar to the one described for \testA\ can be observed, although the shocks are now stronger, because the lower value of the speed of sound (see $c_0$ in Tab.~\ref{t3:state}) leads, given the same piston velocity, to a higher Mach number.
For the same reason, also the wave propagation speed is lower.

The results of the first tests and the previous analysis demonstrate that the proposed method is capable of detecting all expected NICFD phenomena, without introducing spurious oscillations due to mesh adaptation, and it is robust.

\subsection{2D oscillating piston in a finite\hyp{}length tube}
\label{ssec:finite}
\begin{figure}
\includegraphics[width=\hsize]{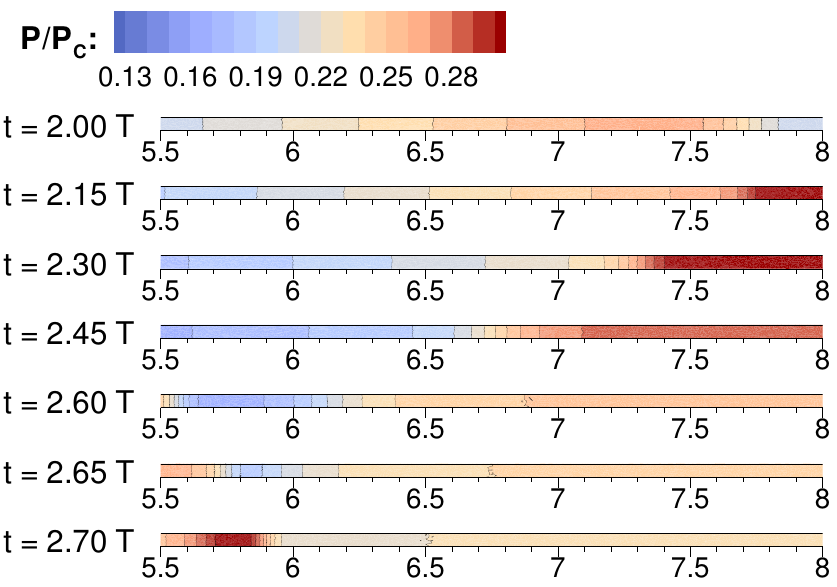}\\[1ex]
\hspace*{2mm} \includegraphics[scale=1]{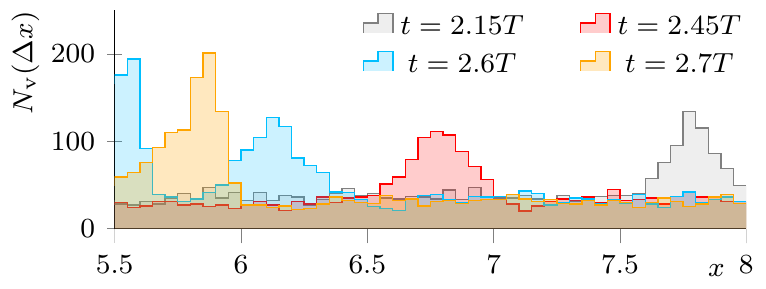}
\caption{Oscillating piston in a finite\hyp{}length tube, \testA: pressure contour plots and histograms of the grid nodes (bin width $\Delta x=0.05$), near the right closed end, during the third oscillation period. 
The first compressive waves originated on the right piston side reach the solid wall after $t=2T$.
The reflected waves interact first with the rarefaction waves and then with the compressive waves, generating a strong shock (approximately around $x=5.9$ at $2.65<t/T<2.7$).
}
\label{fFT:contourDil}
\end{figure}

\begin{figure}
\includegraphics[width=\hsize]{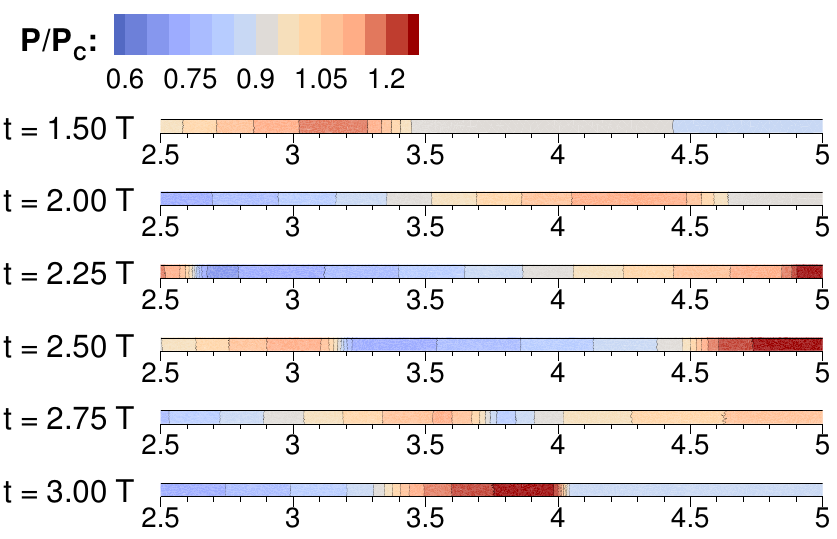}\\[1ex]
\hspace*{2mm} \includegraphics[scale=1]{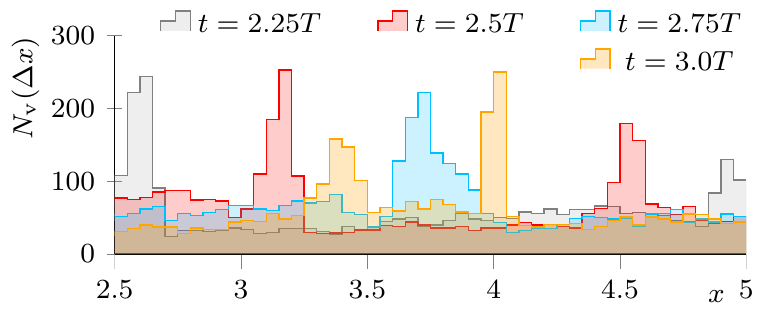}
\caption{Oscillating piston in a finite\hyp{}length tube, \testB: pressure contour plots and histograms of the grid nodes (bin width $\Delta x=0.05$), near the right closed end, during the third oscillation period. 
The first compressive waves originated on the right piston side reach the solid wall after $t=2T$.
The reflected waves interact first with the rarefaction waves and then with the compressive waves, generating a strong shock (approximately around $x=4$ at $2.75<t/T<3$).
}
\label{fFT:contourNICFD}
\end{figure}
%
In the second type of test, presented here, the piston moves inside a finite\hyp{}length tube, to simulate the reflection of the waves by solid walls and the interaction between waves with opposite directions.
The harmonic motion imposed to the piston is the same one described in the previous section.
In \testA, the lengths of the tube are $L_\mathrm{L}=10~\mathrm{m}$ and $L_\mathrm{R}=8~\mathrm{m}$ and the initial grid is composed of $4056$ nodes and $6306$ triangles.
In \testB, as the wave propagation speed is lower, the tube is shorter, $L_\mathrm{L}=8~\mathrm{m}$ and $L_\mathrm{R}=5~\mathrm{m}$, and the initial grid consists in $2927$ nodes and $4548$ elements.

The details of the reflection of the waves at the right wall are shown in Fig.~\ref{fFT:contourDil} and \ref{fFT:contourNICFD}.
When a compressive wave reaches the right wall, it is reflected back and starts moving upstream, in the direction opposite to the flow. Thus, it interacts first with the expansion waves, reducing its intensity, then it comes into contact with the incoming compressive waves. The interaction between compressive waves of opposite family originates two compressions with different strength that move outwards from the interaction point.

Mesh adaptation is crucial to correctly detect such a complex, variable flow pattern, where the a priori determination of the locations of all relevant flow features is not practical.
The histograms of the grid nodes along the $x$-axis in Fig.~\ref{fFT:contourDil} and \ref{fFT:contourNICFD} highlight that the grid density is properly modified according to the position and the intensity of the flow variations.
In particular, the histograms in Fig.~\ref{fFT:contourNICFD} display at times $t=2.25T$ and $t=2.5$ two peaks corresponding to the incoming compressive waves (left) and to the reflected one (right) which move towards each other; as the waves coalesce at $t=2.75T$ only one peak is present; but at $t=3T$ when the compressive waves go in opposite directions, two refined regions are again generated.
So, the present test assesses also the capability of the proposed adaptive method to deal with wave interactions and reflections in the NICFD regime, which is an essential feature to simulate more complex and realistic flow fields.

\subsection{2D impulsively started piston}
\label{ssec:impulsive2d}
\begin{figure*}
\includegraphics[width=\hsize]{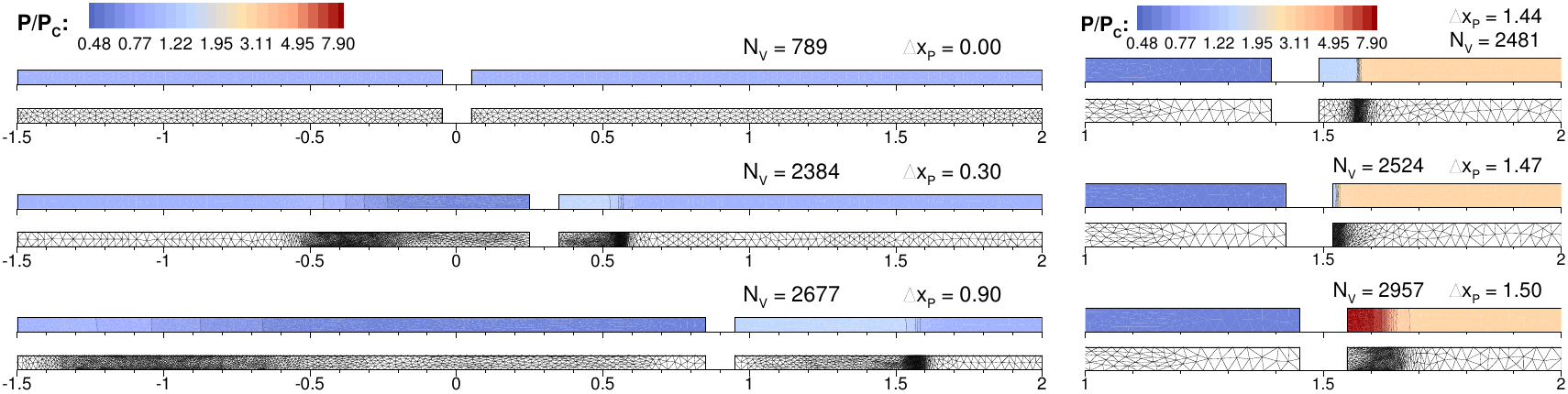} 
\caption{Impulsively started piston in NICFD regime (\testB): pressure contour plots and grids. At left: the whole computational domain at the initial time and at two instants before reflection of the waves. Right: details of the second shock reflection, which occurs on the right piston surface.}
\label{fIS:contour}
\end{figure*}
\begin{figure}
\includegraphics[width=\hsize]{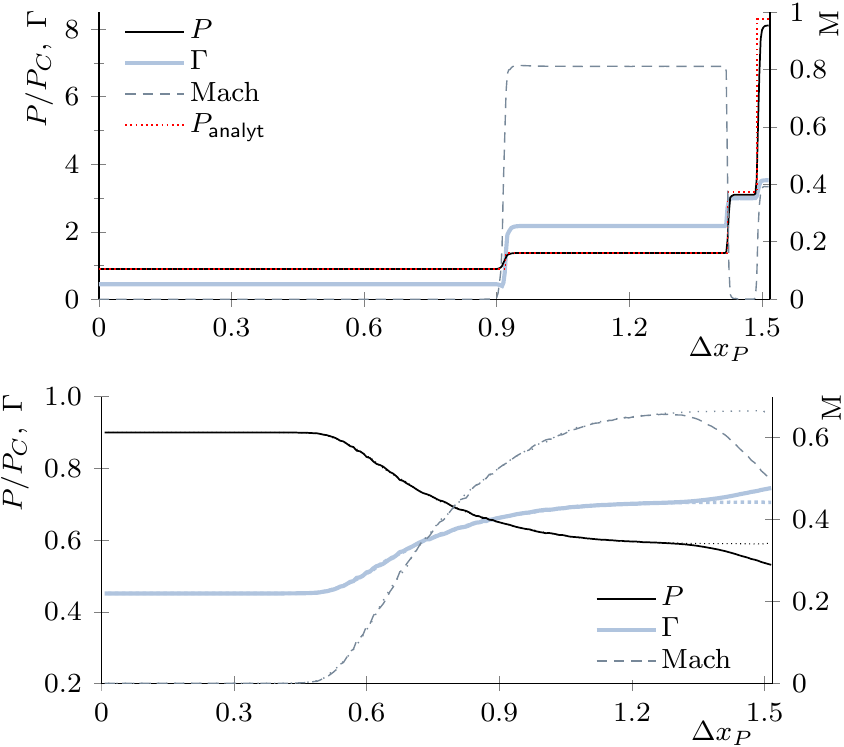}
\caption{Impulsively started piston in NICFD regime (\testB): temporal evolution of pressure (\line{Black, solid, line width=0.5pt}), Mach (\line{mark=none, densely dashed, LightSlateGrey}) and fundamental derivative (\line{LightSteelBlue, solid, line width=1.2pt}) at $\bar{x}=1.6$ (top) and at $\bar{x}=-0.8$ (bottom). In the top plot, the dotted line displays the pressure trend obtained analytically. In the bottom plot, the dotted lines show the evolution of the same quantities in case of an open-end tube, i.e. with no reflection. }
\label{fIS:profile}
\end{figure}

%
The last 2D test case in the NICFD regime is the impulsively started piston in a closed tube, with sizes $L_\mathrm{L}=1.5~\mathrm{m}$ and $L_\mathrm{R}=2~\mathrm{m}$.
For brevity, only \testB\ is shown.
At time $0$, the piston is instantaneously set into motion with a constant velocity $V_\mathrm{P}=47.12~\mathrm{m/s}$, generating a shock wave in the right part of the tube and an expansion fan in the left part.
The simulation is performed until the piston displacement is $\Delta x_\mathrm{P}=1.51~\mathrm{m}$, so that the shock is reflected twice, once by the right end of the tube and once by the piston face.
The initial grid is composed of $789$ nodes and $1574$ elements
and the whole simulation time is divided in $N_T=1010$ time steps.
On the smallest allowed edge ($h_\mathrm{min}=0.008~\mathrm{m}$), the Courant number is $\mathrm{Co}=1.8$ 
with respect to the piston velocity, but $\mathrm{Co}=3.1$ with respect to the velocity of the first generated shock.

Figure~\ref{fIS:contour} displays the results of the flow field before the reflections of the shock and the expansion fan, and after the shock reflection on the piston surface. In these pictures the effectiveness of mesh adaptation is evident, especially after the second reflection, when the strongest shock occurs. 
A more quantitative view of the results is given by Fig.~\ref{fIS:profile}, which shows the pressure, Mach, and fundamental derivative profiles at two constant locations, in the left and right part of the tube.
From the evolution of the fundamental derivative, we can say that the left part of the tube remains for the whole test in the NICFD region, $0<\Gamma<1$, while the right part ends in the supercritical regime. 
For the location near the left end of the tube, the results of the open-end case are also shown to highlight the velocity decrease imposed by the no-penetration boundary condition.
For the location near the right end, the pressure evolution is computed also analytically according to the shock conditions. A fairly good agreement between the analytical and the numerical result is obtained, 
although a deviation between the pressure values occurs after the second reflection and becomes larger with the third reflection.

This test proves the capability of the proposed adaptive scheme to deal with sudden generation of expansion fans and shock waves in the NICFD regime, as well as with their interaction with solid boundaries.
The specification of the maximum tolerated error during the metric construction allows stronger or weaker grid refinement to be achieved according to the strength of the flow phenomena without requiring any user's action.

\subsection{2D oscillating piston in non-classical region}
\label{ssec:nonClass}
\begin{figure}
  \includegraphics[width=\hsize]{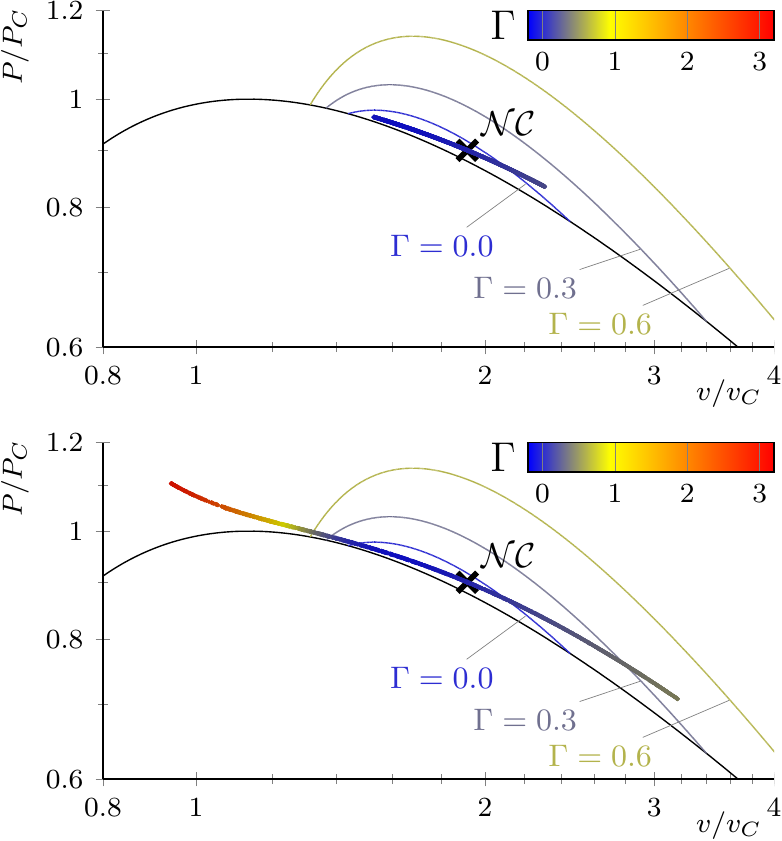}
\caption{Oscillating piston in non-classical region, \testC: results of the motion with small (top) and large amplitude (bottom).
The flow fields resulting after one period of oscillation in the infinite-length tube are plotted and colored according to the value of $\Gamma$, by using the same color map.
Pressure and specific volumes are scaled with respect to the critical values reported in Tab.~\ref{t2:md4m}.
}
\label{fNC:VLE}       
\end{figure}
\begin{figure*}
  \includegraphics[width=\hsize]{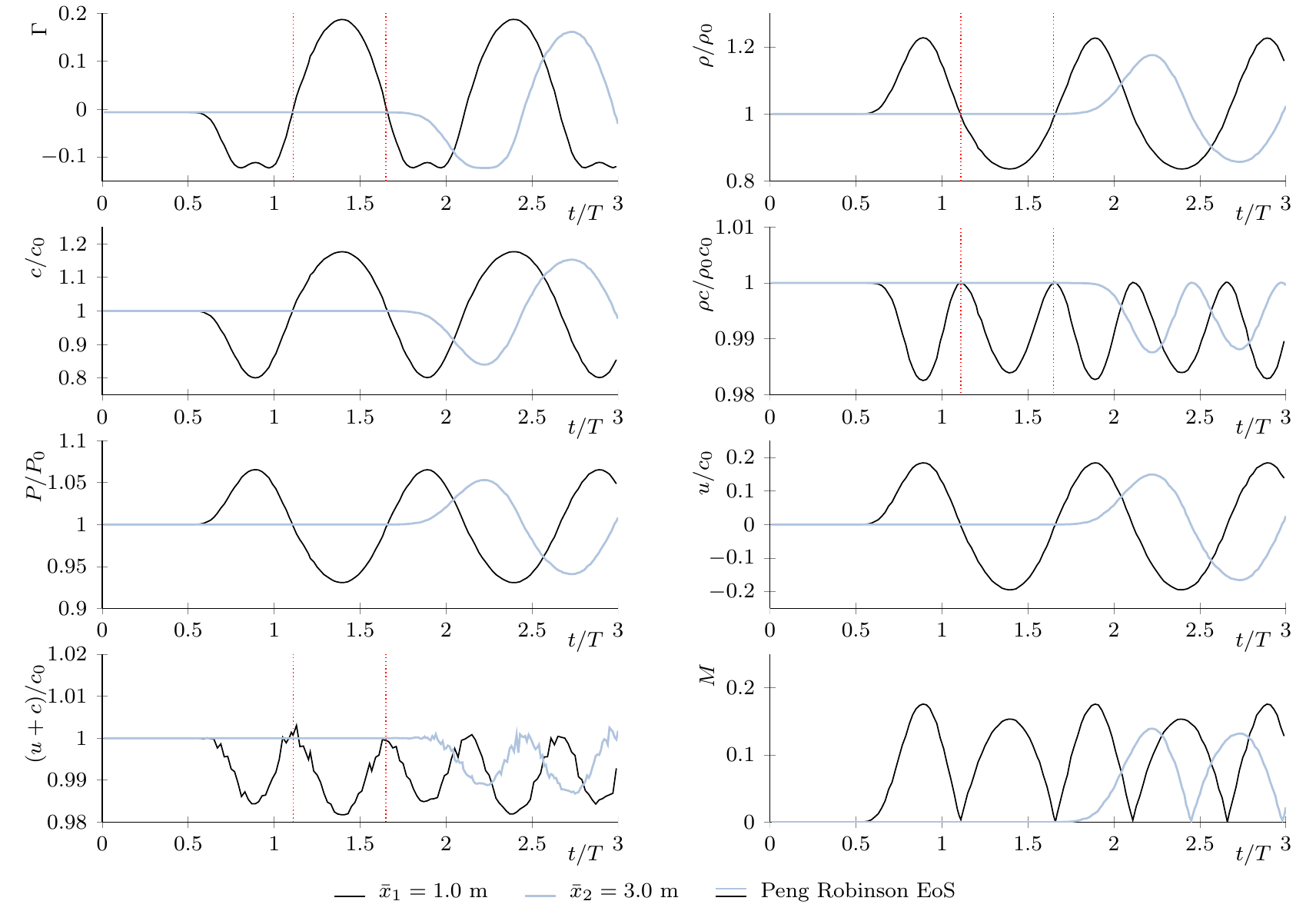}
\caption{Oscillating piston in non-classical region, \testC, small amplitude: temporal evolution of eight thermodynamic quantities at two different locations in the right part of the tube, one near the piston ($\bar{x}_1$, dark lines~\line{black}) and one farther ($\bar{x}_2$, light\hyp{}colored lines~\line{LightSteelBlue}). From the first row to the last one: fundamental derivative, density, speed of sound, acoustic impedance, pressure, $x$-component of the fluid velocity, wave propagation speed, and Mach number. 
Pressure, density, and velocities are scaled with respect to the initial values $P_0$, $\rho_0$, and $c_0$ reported in Tab.~\ref{t3:state}.
For $\bar{x}_1$, the vertical dotted lines (\line{densely dotted, red}) highlight the first two points where $\Gamma=0$.
}
\label{fNC:probeSmall}       
\end{figure*}
\begin{figure*}
  \includegraphics[width=\hsize]{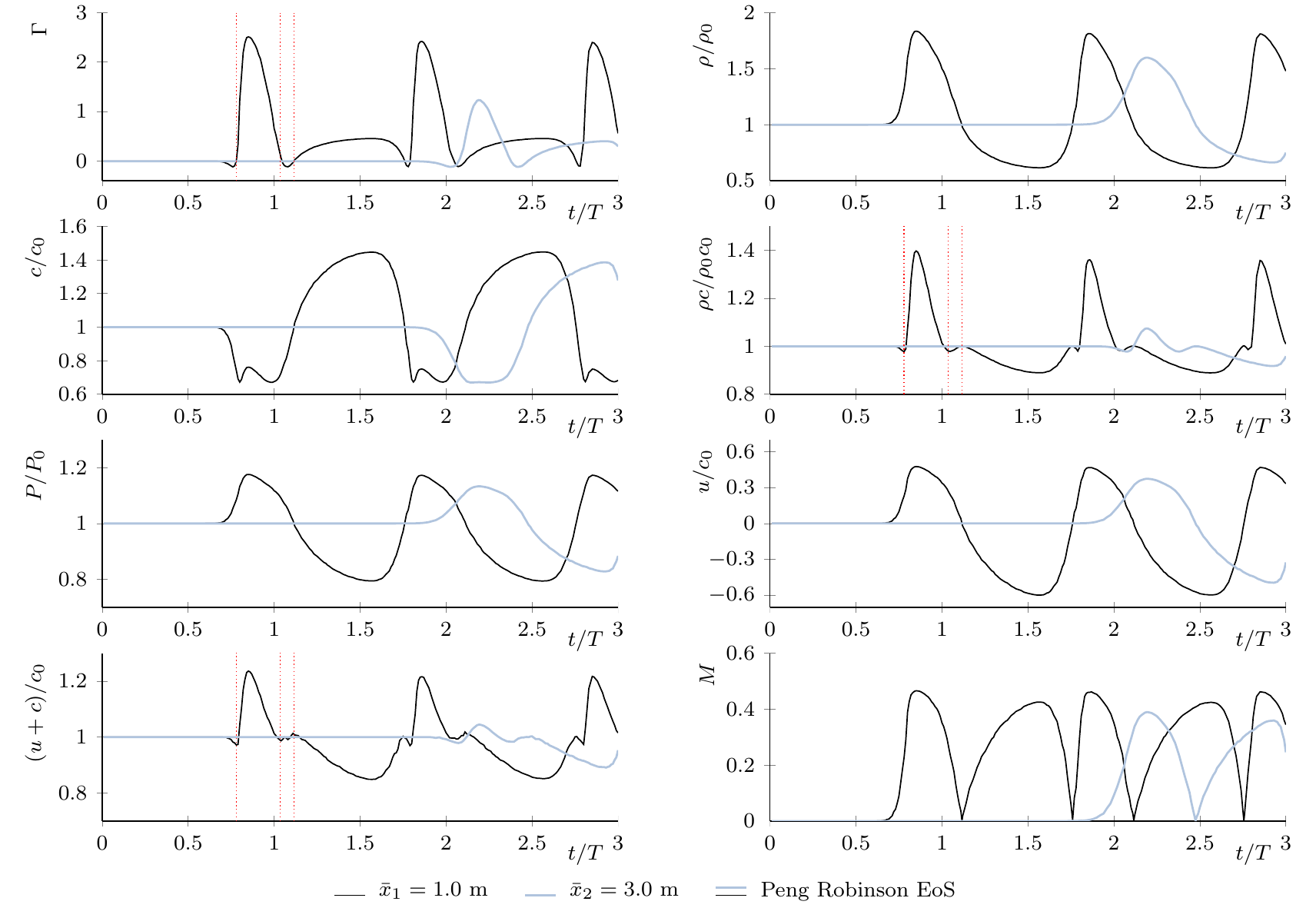}
\caption{Oscillating piston in non-classical region, \testC, large amplitude: temporal evolution of eight thermodynamic quantities at two different locations in the right part of the tube, one near the piston ($\bar{x}_1$, dark lines~\line{black}) and one farther ($\bar{x}_2$, light\hyp{}colored lines~\line{LightSteelBlue}). From the first row to the last one: fundamental derivative, density, speed of sound, acoustic impedance, pressure, $x$-component of the fluid velocity, wave propagation speed, and Mach number. 
Pressure, density, and velocities are scaled with respect to the initial values $P_0$, $\rho_0$, and $c_0$ reported in Tab.~\ref{t3:state}.
For $\bar{x}_1$, the vertical dotted lines (\line{densely dotted, red}) highlight the first three points where $\Gamma=0$.
}
\label{fNC:probeLarge}       
\end{figure*}
%
A couple of tests inside and across the non\hyp{}classical region, at $\Gamma<0$, are now presented.
The piston oscillates harmonically inside an infinite\hyp{}length tube, but with respect to the previous test cases, the oscillation amplitude is now reduced.
More specifically, a small amplitude $A=-0.05~\mathrm{m}$ is selected to constrain the fluid state to lie inside or near the region at $\Gamma<0$, and a large amplitude $A=-0.15~\mathrm{m}$ is tested to generate stronger perturbations.
The evolution of the flow field is shown in the $P$--$v$ plane in Fig.~\ref{fNC:VLE}. The flow state evolves in close proximity to the VLE curve, being the initial state on an isotherm below the critical one ($T_0 = 0.996 T_c$).

For the motion characterized by the small amplitude, the evolution of different thermodynamic quantities at two fixed locations is displayed in Fig.~\ref{fNC:probeSmall}.
Since this motion does not introduce strong perturbations, the flow field on the right side of the tube evolves approximately along an isentrope.
During the first part of the compression, the flow state moves more inside the non\hyp{}classical region (see also Fig.~\ref{fNC:VLE}), reaching a minimum  in the value of the fundamental derivative, but during the final part of the compression, $\Gamma$ slightly increases, as the flow is leaving the core of the non\hyp{}classical region. From here, the successive expansion crosses the core of the non\hyp{}classical region and ends in the NICFD region, reaching the maximum value of $\Gamma$ (equal to 0.187 for the location $\bar{x_1}$), from which the next compression starts.  
Comparing to the NICFD results of Sec.~\ref{ssec:infinite2D}, the most notable differences concern the acoustic impedance and the wave propagation speed.
Indeed, re\hyp{}writing the fundamental derivative as $\Gamma=\frac{1}{c}\left(\frac{\partial \rho c}{\partial \rho}\right)_s$, if an isentropic expansion starts in the non\hyp{}classical region and evolves outwards, the acoustic impedance has a maximum at $\Gamma=0$.  
In this specific test, since the initial value of $\Gamma$ is close to $0$, the maximum of the acoustic impedance is extremely close to the initial value, and $\rho c < \rho_0 c_0$ everywhere.
A similar evolution can be observed also for the wave speed, which remains almost everywhere smaller, though close, to the initial value.
According to Eq.~\eqref{e:duc}, $\mathrm{d}(u+c)$ is positive in the first part of the expansion ($\mathrm{d}P<0$ and $\Gamma<0$) and negative in the second, and vice versa for the compression. Observing that in this test all pressure maxima occurs at $\Gamma<0$ and the minima at $\Gamma>0$, the wave propagation speed has minima when $\mathrm{d}P=0$ and maxima when $\Gamma=0$.
It is remarkable that such an ``acoustic'' behavior is obtained for finite amplitude waves because the value of $\Gamma$ is close to zero~\cite{Thompson1972}.

Fig.~\ref{fNC:probeLarge} shows the same quantities for the test in which the large oscillation amplitude is imposed.
In this test, the flow state already exits the non\hyp{}classical region during the first compression and most part of the motion occurs in the NICFD region. Therefore, both non\hyp{}ideal and non\hyp{}classical phenomena occur.
We can observe the non\hyp{}monotone variation of speed of sound during the isentropic expansion, as seen for \testB\ in Sec.~\ref{ssec:infinite2D}.
As highlighted for the small\hyp{}amplitude test, the acoustic impedance has minima and maxima for $\Gamma=0$ (see the vertical dotted lines in Fig.~\ref{fNC:probeLarge}), but these are not the absolute extrema, which occur in correspondence of the extrema of $\rho$ and $c$.
Similarly, the wave propagation speed exhibits local minima and maxima for $\Gamma=0$, but the absolute ones reflect the extrema of the pressure, since all points at which $\mathrm{d}P=0$ fall in the region at $\Gamma>0$.

Also in these two tests, the proposed mesh adaptation strategy allows to automatically modify the grid spacing to increase the solution accuracy, without introducing oscillations due to solution interpolation.
The importance of this feature can be appreciated by looking at Fig.~\ref{fNC:VLE}: some points of the flow field are very close to the VLE curve and oscillations may cause them to enter the two-phase region, where the speed of sound may become negative and ad hoc numerical tools are required for the thermodynamic modeling.

\subsection{3D assessment in the NICFD regime}
\label{ssec:results3d}
\begin{figure}
\centering
\includegraphics[width=\hsize]{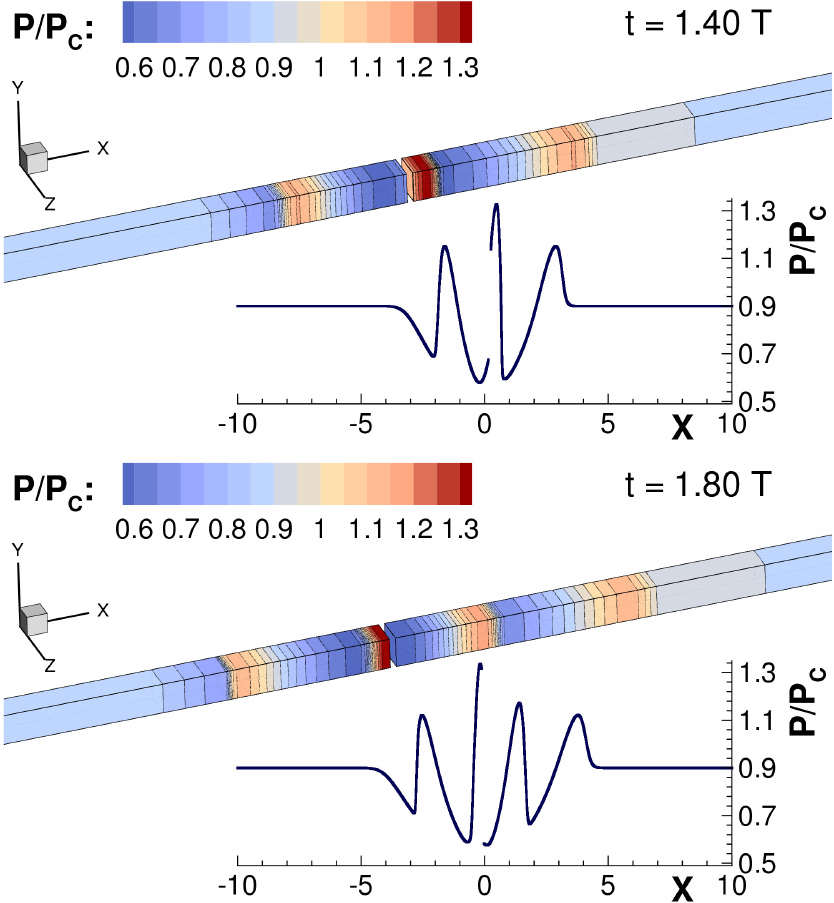}
\caption{Oscillating piston in an infinite\hyp{}length tube, \testB, 3D: pressure contour plot and pressure profile at two different times during the second oscillation period. Pressure is scaled with respect to the critical value.
To make clearer the 3D picture, the $y$ and $z$ axis have been multiplied by a factor 10 with respect to the $x$ axis.
In the pressure profile, each point corresponds to a grid node.
}
\label{f3d:oscill_contour}
\end{figure}

In this final subsection, we briefly present some 3D results to assess the proposed adaptive scheme also for tetrahedral grids.
First, we reproduce an oscillating piston in an infinite\hyp{}length tube, the \testB\ of Sec.~\ref{ssec:infinite2D}.
The initial grid is made of $27806$ nodes and $97259$ tetrahedra.
Given that the physical problem is one-dimensional, we can make a comparison between the results obtained in the 2D and 3D simulations.
Figure~\ref{f3d:oscill_contour} shows the pressure fields at two different times during the second oscillation period, i.e., at $t=1.4~T$ and $t=1.8~T$, which are the same shown in Fig.~\ref{fOB:cflNICFD}.
A good matching between the 3D and 2D pressure profiles is obtained.
Furthermore, as observed already in the 2D tests, the profiles in Fig.~\ref{f3d:oscill_contour} are in fact scatter plots of all grid points, and the method correctly reproduces the physical one\hyp{}dimensional behavior of the flow assessing the validity of the proposed approach.

\begin{figure*}
\includegraphics[width=\hsize]{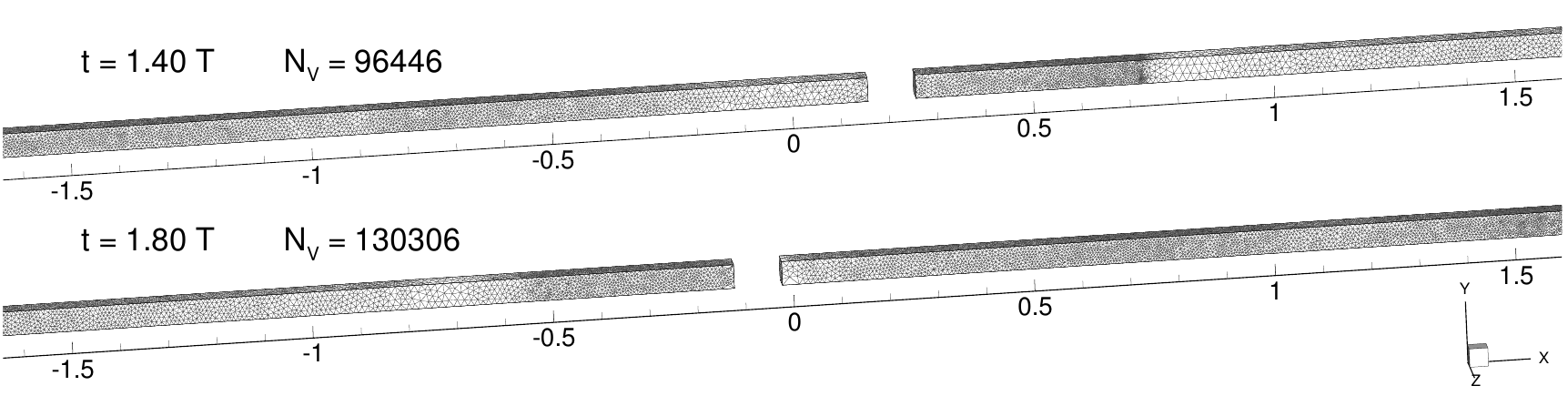}
\caption{Oscillating piston in an infinite\hyp{}length tube, \testB, 3D: detail of the grid at two different times during the second oscillation period, at $t=1.4 T$ and $t=1.8T$. Watching also at the pressure fields shown in Fig.~\ref{f3d:oscill_contour}, we can notice how the grids are modified according to the solution behavior.}
\label{f3d:oscill_grid}
\end{figure*}

In this test, a target isotropic metric is defined by combining an adaptation criterion based on the Hessian of the Mach number and another criterion based on the gradient of the pressure.
A portion of the resulting grids at the previous times is shown in Fig.~\ref{f3d:oscill_grid} to highlight the crucial role of mesh adaptation also in 3D.
By comparing them with the pressure fields shown in Fig.~\ref{f3d:oscill_contour}, the grids result to be refined near the sharper variations and coarse in quasi\hyp{}uniform regions.

\begin{figure*}\sidecaption
\includegraphics[scale=1]{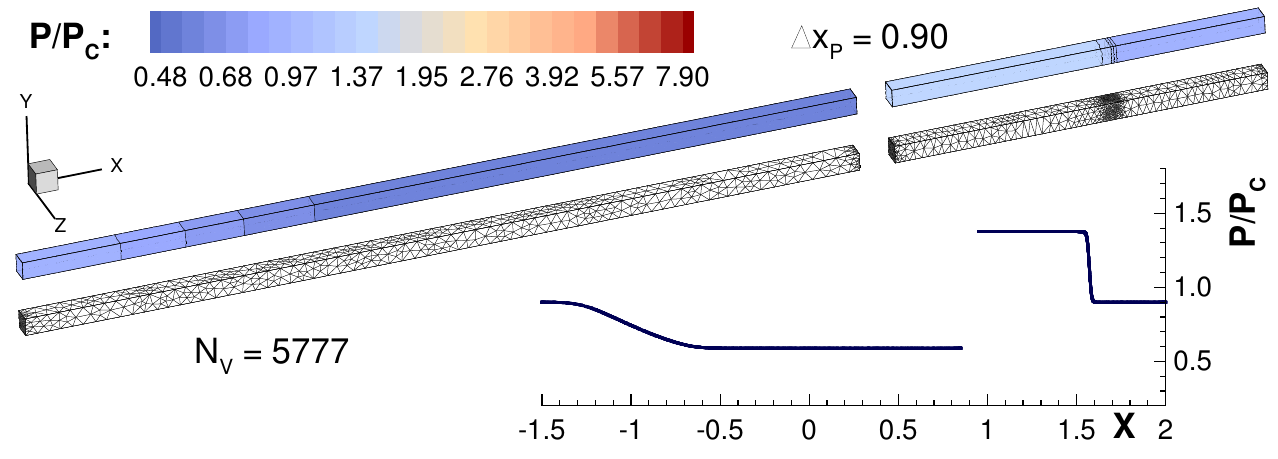}
\caption{Impulsively started piston in NICFD regime (\testB), 3D: pressure contour plot, grid and pressure profile, when the piston displacement is $\Delta x_\mathrm{P}=0.9$, that is before any wave reflection.
}
\label{f3d:impulsive1}
\end{figure*}

\begin{figure*}\sidecaption
\includegraphics[scale=1]{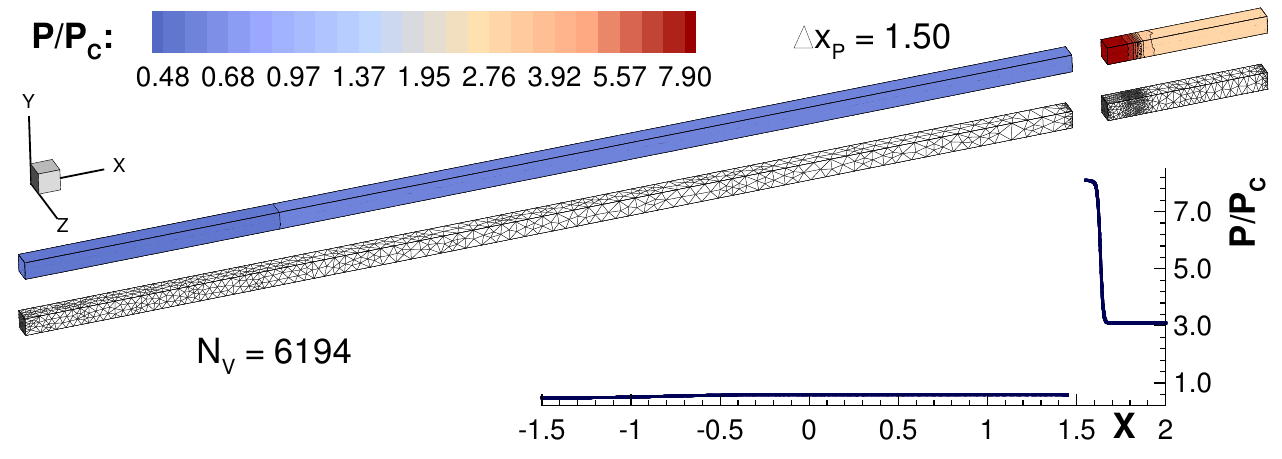}
\caption{Impulsively started piston in NICFD regime (\testB), 3D: pressure contour plot, grid and pressure profile, when the piston displacement is $\Delta x_\mathrm{P}=1.5$, that is after the expansion fan has been reflected by the left wall and shock waves have been reflected by the right piston wall (second shock reflection).
}
\label{f3d:impulsive2}
\end{figure*}

Lastly, a test case of the impulsively started piston is performed. We simulate the same physical features described in Sec.~\ref{ssec:impulsive2d} to be able to compare 2D and 3D results.
As in the previous test, three-dimensional isotropic mesh adaptation is performed by exploiting a compound error estimator that includes the Hessian of the Mach number and the gradient of the pressure.
The pressure field and the computational grid are shown in Fig.~\ref{f3d:impulsive1} before any wave interaction with the solid wall, and in Fig.~\ref{f3d:impulsive2} after the reflection of the expansion fan on the left side of the tube and the second shock reflection, which occurs on the piston right face.
The effect of mesh adaptation is clear, especially after the reflections of the expansion fan and of the shock. In particular, we can observe the grid portion close to the shock reflected by the right piston wall, which is the strongest phenomenon in the flow field, and the one next to the left end of the tube, where the rarefaction fan is reflected. Because of the different intensity of the phenomena, a different level of refinement is automatically performed, i.e., a smaller grid spacing is reached near the shock, without the user's intervention.

Although concise, these results show that the proposed method works well also for three\hyp{}dimensional simulations. From the displayed contour plots and pressure profiles, we can notice that no spurious oscillations are introduced by the interpolation\hyp{}free adaptation and the physical mono\hyp{}dimensional behavior is  achieved.

\section{Conclusions}
\label{sec:concl}
This paper presents and discusses the assessment of mesh adaptation in unsteady NICFD simulations, since a detailed investigation of the possible extension of standard adaptation techniques, widely used in ideal gas simulations to deal with geometrically complex moving\hyp{}body problems or to improve the solution accuracy, to such a peculiar thermodynamic regime was missing.
Among the available adaptation techniques, we have selected an interpolation\hyp{}free adaptive scheme capable of representing the volume changes due to mesh adaptation as fictitious continuous deformations of the finite volumes that compose the domain, and of treating them within the ALE framework.
In our opinion, this approach represents an advantage for mesh adaptation in the NICFD regime since it prevents the introduction of spurious oscillations due to interpolation, which may unpredictably modify the thermodynamic state of the flow, moving it under the VLE curve.

The proposed method has been assessed through two sets of numerical tests. First, steady simulations of oblique and conical shocks were performed in the unsteady fashion to assess the validity of the numerical scheme. Then, different problems involving a piston moving into a tube filled with \mdqm\ were simulated. Two initial conditions have been selected along an isotherm above the critical one: \testA\ is representative of the dilute regime (approximately $0.9<\Gamma<1$) and \testB\ is more in the core of the NICFD regime.
A third initial state has been chosen near the VLE curve, along an isotherm below the critical one, to perform tests across the non-classical and non-ideal region.
The thermodynamic behavior of the flow has been described by the polytropic Peng\hyp{}Robinson EoS.
Different simulations of the harmonic motion and the impulsive starting of the piston were performed, both in an infinite\hyp{}length and in a closed tube.

The expected NICFD and non\hyp{}classical behavior was well reproduced by the numerical results and mesh adaptation techniques have been successfully exploited to adapt the computational grid to the unsteady solution.
In particular, in all tests, the resulting grid was refined near shocks, rarefaction and expansion waves, and their evolution in the flow field was accurately tracked, even if large deformations were imposed on the domain boundaries.
At the same time, the adaptation strategy was capable of removing grid nodes where the solution was smooth, without deteriorating the solution accuracy.
Also the interactions between waves and solid walls, or other waves, were correctly treated. 
No spurious oscillations appeared in the results, despite quite large time steps having been enforced.

In conclusion,
the proposed method was assessed to perform adaptive unsteady NICFD simulations.
Therefore, in future works it could be successfully exploited to perform more complex unsteady simulations.
For instance, it could be useful to devise new experimental test rigs to investigate thermo\hyp{}physical properties of fluid flows in the NICFD regime, or to improve the design of the components of systems operating in the NICFD regime, as Organic Rankine Cycles or supercritical carbon dioxide power cycles.

\begin{acknowledgements}
This study was partially funded by the European Research Council (Consolidator Grant N. 617603, Project
NSHOCK, funded under the FP7-IDEAS-ERC scheme).
\end{acknowledgements}

\bibliographystyle{spmpsci_unsort}     
\bibliography{biblioShockWaves}   

\end{document}